\documentclass[12pt,preprint]{aastex}

\RequirePackage{lineno} 
\usepackage{longtable}
\usepackage{epstopdf}  
\usepackage{ifthen}
\usepackage{rotating}
\usepackage{subfigure} 
\usepackage{booktabs}

\def\afrho{{$A$($\theta$)$f\rho$}}
\def\deg{$^\circ$}
\def\epoxi{{\it EPOXI}}
\def\icarus{Icarus}
\def\rh{$r_\mathrm{H}$}

\begin{document}
\bibliographystyle{icarus}

\setlength{\footskip}{0pt} 

\title{The Highly Unusual Outgassing of Comet 103P/Hartley 2 from Narrowband Photometry and Imaging of the Coma}

\author{Matthew M. Knight \altaffilmark{1,2,3} and David G. Schleicher \altaffilmark{2}}

\altaffiltext{1}{Contacting author: knight@lowell.edu.}
\altaffiltext{2}{Lowell Observatory, 1400 W. Mars Hill Rd, Flagstaff, AZ 86001, U.S.A.}
\altaffiltext{3}{The Johns Hopkins University Applied Physics Laboratory, 11100 Johns Hopkins Road, Laurel, Maryland 20723}




\begin{singlespace}

\section*{Abstract}
We report on photometry and imaging of Comet 103P/Hartley 2 obtained at Lowell Observatory from 1991 through 2011. We acquired photoelectric photometry on two nights in 1991, four nights in 1997/98, and 13 nights in 2010/11. We observed a strong secular decrease in water and all other observed species production in 2010/11 from the 1991 and 1997/98 levels. We see evidence for a strong asymmetry with respect to perihelion in the production rates of our usual bandpasses, with peak production occurring $\sim$10 days post-perihelion and production rates considerably higher post-perihelion. The composition was ``typical,'' in agreement with the findings of other investigators. We obtained imaging on 39 nights from 2010 July until 2011 January. We find that, after accounting for their varying parentage and lifetimes, the C$_2$ and C$_3$ coma morphology resemble the CN morphology we reported previously. These species exhibited an hourglass shape in October and November, and the morphology changed with rotation and evolved over time. The OH and NH coma morphology showed hints of an hourglass shape near the nucleus, but was also enhanced in the anti-sunward hemisphere. This tailward brightness enhancement did not vary significantly with rotation and evolved with the viewing geometry. We conclude that all five gas species likely originate from the same source regions on the nucleus, but that OH and NH were derived from small grains of water and ammonia ice that survived long enough to be affected by radiation pressure and driven in the anti-sunward direction. We detected the faint, sunward facing dust jet reported by other authors, and did not detect a corresponding gas feature. This jet varied little during a night but exhibited some variations from night to night, suggesting it is located near the total angular momentum vector. Overall, our imaging results support the conclusions of other authors that Hartley 2's ``hyperactivity'' is caused by icy particles of various sizes that are lifted off the surface and break up in the coma to greatly increase the effective active surface area.

\begin{description}
\item{\textbf{Keywords}:} Comets; Comets, coma; Comets, composition
\end{description}

\section{INTRODUCTION}
Jupiter-family comet 103P/Hartley 2 became one of the best-studied comets of all time when, in late-2010, it simultaneously passed very close to the Earth and was the subject of a flyby by the {\it EPOXI} spacecraft (e.g., \citealt{meech11}). First results have already been published for {\it EPOXI} spacecraft data \citep{ahearn11a}, radar observations of the nucleus \citep{harmon11}, 
composition of the coma \citep{combi11b,dellorusso11,drahus11,mumma11,weaver11,hartogh11}, and morphology of the coma \citep{knight11b,samarasinha11,lara11,waniak12}. These and earlier results will undoubtedly be discussed in more detail elsewhere in this Special Issue; in the interest of brevity, we restrict our discusion to those results having bearing on our current work.

We obtained both photometry and imaging of Hartley 2, and discuss the observations and reductions in Section~\ref{sec:obs}. In Section~\ref{sec:photometry} we discuss our photoelectric photometer measurements acquired during the 1991, 1997/98, and 2010/11 apparitions. Due to unfavorable viewing geometry, we did not observe Hartley 2 on its two other known apparition, 1986 (when it was discovered) or 2004. Previous investigators \citep{weaver94,ahearn95} found a typical composition and a low dust-to-gas ratio. Assuming that Hartley 2 had a comparable active fraction to other comets ($\sim$10\%), it was inferred to have a rather large nucleus (of order a few km). The \epoxi\ flyby conclusively showed that Hartley 2 has a small nucleus (effective radius of 0.57$\pm$0.02 km; \citealt{ahearn11a}), confirming the conclusions of remote studies of the nucleus \citep{groussin04,lisse09}. \epoxi\ also revealed that the active fraction of the nucleus is relatively low and thus its ``hyperactivity'' necessitates an extended source of volatiles. Numerous new water production measurements have recently been published in support of \epoxi\ and are compared with our measurements.

We also follow up our previous investigation of the CN coma morphology in the 2010/2011 apparition. We discovered the existence of time-varying CN coma structures and used the repetition of these structures to measure a rotation period \citep{iauc9163a}. In \citet{knight11b}, henceforth ``Paper 1'', we interpreted the morphology as likely being due to two ``jets'' whose appearance varied as the nucleus rotated, and evolved during the apparition. We used repetitions in the morphology to measure rotation periods monthly from 2010 August through 2010 November and determined that the rotation period was increasing during this time from $\sim$16.7 hr in August to $\sim$18.7 hr in November. We noted that the morphology did not exactly repeat from cycle to cycle, but repeated better after 3, 6, 9... cycles and inferred that it was in non-principal axis rotation with a component period roughly three times as long as the measured rotation period. Similar coma morphology and rotation periods were reported by \citet{samarasinha11}, whose images covered a comparable range of dates (2010 September 1 through December 15). \citet{lara11} acquired snapshot imaging near perigee, reporting similar CN morphology and showing that the C$_3$ coma morphology also resembled the CN coma morphology. Lara {\it et al.} also reported the existence of a much smaller-scale dust jet in the sunward direction. A similar feature is visible in movies made from \epoxi\ HRI images posted on the \epoxi\ website\footnote{http://epoxi.umd.edu/}. \citet{waniak12} obtained CN and C$_3$ images near the time of the \epoxi\ flyby, finding a similar periodicity as we reported and noting the agreement of the CN and C$_3$ morphology. In Section~\ref{sec:gas_morphology} we intercompare the coma morphology exhibited by various gas species (CN, C$_3$, C$_2$, OH, and NH) in 2010 October and November, and discuss the dust morphology, in particular looking for evidence of the dust jet. 

Finally, in Section~\ref{sec:disc} we discuss our results and how they fit into the larger body of knowledge about Hartley 2 acquired by \epoxi\ and numerous remote observers. We summarize our findings in Section~\ref{sec:summary} and briefly discuss their application to future modeling efforts. 

\section{OBSERVATIONS AND REDUCTIONS}
\label{sec:obs}
\subsection{Instrumentation}
The goals of our studies of Hartley 2 have varied, with composition and production rate studies the emphasis of the 1991 and 1997/98 apparitions and multiwavelength imaging the emphasis in 2010/11 (with production rate and compositional monitoring a secondary emphasis). Thus, our techniques and instrumentation have varied. The production rate and composition studies continue to be made with a traditional photoelectric photometer. This provides superior signal-to-noise for bulk coma measurements as compared to a CCD and also ensures continuity with earlier data sets. The imaging necessarily requires use of a CCD. All photometer observations were obtained with the Hall 42-in (1.1-m) telescope at Lowell Observatory except for one night in 1997 which was made using Lowell's 0.9-m telescope at Perth Observatory. Imaging observations were acquired with either the Hall 42-in or the 31-in (0.8-m) telescopes at Lowell Observatory. In total, we acquired two nights of photometry in 1991, four nights of photometry in 1997/98, 13 nights of photometry in 2010/11, and 39 nights of imaging in 2010/11. The observing circumstances for all nights of imaging and photometry are given in Tables~\ref{t:imaging_circ} and \ref{t:phot_circ}, respectively.



The photometers used had EMI 6256 photomultiplier tubes
and pulse counting systems except for the sole night (1997 December 1)
at Perth Observatory where a DC amplifier was employed.
The Hale-Bopp (HB) narrowband filter set \citep{farnham00} was utilized in 1997 and later,
superceding the International Halley Watch (IHW) set \citep{ahearn91} used in 1991.
HB filters were also used for the CCD observations, along with a
broadband Cousins R filter for increased signal-to-noise for dust measurements.
The narrowband comet filters used for both our photometry and imaging isolate emission from daughter gas species (OH, NH, CN, C$_3$, C$_2$) or reflected solar continuum from dust (continuum filters in the UV (UC), blue (BC), green (GC), and red (RC)). 
The comet filters are regularly used to study dust and coma in comets (e.g., \citealt{woodney02,schleicher06a,farnham07a,lederer09,waniak09}). Using the method described in \citet{ahearn95} and \citet{farnham00}, we ``decontaminate'' the data by removing the underlying signal due to reflected solar continuum (and in the case of CN and NH, contamination from C$_3$ as well) from gas bandpasses to yield pure gas and pure dust data. 

The 1.1-m images were obtained using an e2v CCD231-84 chip with 4k$\times$4k pixels which was binned 2$\times$2 at the telescope, resulting in a pixel scale of 0.74 arcsec. The 0.8-m images were obtained with an e2v CCD42-40 chip with 2k$\times$2k pixels and a pixel scale of 0.46 arcsec. At times images were rebinned by an additional factor of 2 during processing to improve the effective signal-to-noise; the final pixel scales of these images were 1.48 arcse for 1.1-m images and 0.92 arcsec for 0.8-m images.

\subsection{CCD Observations and Reductions}
\label{sec:ccd_reductions}
Earlier studies of Hartley 2 had concluded that it has a small, ``hyperactive'' nucleus (e.g., \citealt{groussin04,lisse09}), so our imaging program focused on coma morphology rather than direct nucleus studies. Throughout the apparition, we regularly monitored the comet in broadband R and narrowband CN filters, with occasional narrowband blue continuum (BC) images as well. On photometric nights we used additional HB narrowband filters, increasing the variety and frequency as the comet brightened and shorter exposure times were possible. Images using an H$_2$O$^+$ filter were only obtained on 2010 November 4, in support of concurrent {\it Chandra} observations being conducted by \citet{lisse12}; they are not discussed further in this paper.
See Table~\ref{t:imaging_circ} for a list of the filters used on a given night. Exposure times varied by filter and with the comet's brightness, but ranged from as long as 900 s (OH in September) to as short as 30 s (R in November). Most R, OH, and CN images were obtained in sets of 3--10 images while most images in other filters were obtained as single images. All 1.1-m images were guided at the comet's rate of motion while all 0.8-m images were tracked at the comet's rate of motion (the 0.8-m does not have guiding capability). The HB filters are parfocal and since the primary goal of the imaging campaign was gas coma morphology, we focused for these filters rather than for the R-band. 

Our entire Hartley 2 campaign coincided with a separate effort to measure the nucleus lightcurve of 10P/Tempel 2 \citep{knight12}. As Tempel 2's period is known precisely, and is near 9 hr and changing by $\sim$+0.005 hr per orbit \citep{knight11a}, we required regular monitoring of its lightcurve over many months to detect a change in the period. Fortunately, Tempel 2 generally set as Hartley 2 was rising, but occasionally we split time on both objects. Thus, our temporal coverage of Hartley 2 was slightly less complete than it might have otherwise been, but this did not significantly affect any science results on Hartley 2.

We removed the bias and applied a flat field following standard reduction techniques. We observed HB narrowband standard stars \citep{farnham00} on all photometric nights. For these nights, we followed our standard photometric procedures \citep{farnham00} to determine flux calibrations and process the narrowband images into pure gas and pure dust images. As discussed in Paper 1, there was not significant contamination from the dust continuum in CN images. This allowed us to make morphological assessments of CN images on non-photometric nights (when we could not generate fully ``decontaminated'' images). We normally only observe additional species on photometric nights, but obtained images in other filters despite non-photometric conditions on November 4 in support of the \epoxi\ flyby. 

We centroided by fitting a two-dimensional Gaussian to the apparent photocenter of each image. As discussed in Paper 1, the jets observed in Hartley 2 can pull the centroid away from the nucleus. However, the central condensation generally dominated over any coma features and we estimate that our centroids are accurate to better than 1.5 arcsec. We investigated various enhancement techniques (e.g., \citealt{schleicher03a}) but utilized division by an azimuthal median profile for all images shown in this paper.

\subsection{Photometer Observations and Reductions}
\label{sec:phot_obs}
An observational set for photometry consisted of five gas
filters (OH, NH, CN, C$_3$, and C$_2$) along with two (3650 and 4845 \AA) or
three (3448, 4450, and 5260 \AA) continuum filters associated with the IHW
or HB filter sets, respectively. Three observational sets were obtained
in 1991, four sets in 1997/98, and 55 sets in 2010/11 in support of \epoxi, 
with multiple sets often obtained on a single night.
Photometer entrance aperture diameters ranged between 24 and 156 arcsec
while projected aperture radii ranged from as small as 1950 km near
perigee in 2010 October to as large as 47,900 km, with a median radius
of 10,500 km.

Our standard procedures were used in data acquisition and basic
reductions (see \citealt{ahearn95}), with appropriate changes
associated with the newer HB filters (cf. \citealt{farnham00}); and
improved decontamination for the IHW filters (cf. \citealt{farnham05}).
Gas fluorescence efficiencies that vary with heliocentric velocity are
listed in Table~\ref{t:phot_circ}; sources are given in \citet{schleicher11}.
To extrapolate the resulting column abundances to total coma abundances,
a standard Haser model was applied, using the scalelengths given in
\citet{ahearn95}, and then gas production rates ($Q$) were computed
by dividing the total abundances by the assumed daughter lifetimes.
Finally, water production rates were computed from the OH results
using our empirical conversion (\citealt{cochran93}; also see
\citealt{schleicher98}).
To quantify the dust, we continue to use the quantity \afrho\ 
and, as appropriate, we have now applied a phase correction to the phase angle, $\theta$, listed in Table~\ref{t:phot_circ}
(see \citealt{schleicher11}) to obtain $A$(0{\deg})$f\rho$.
One-sigma uncertainties based on the photon statistics associated
with each data point were computed for the resulting $Q$s and {\afrho}s.
These were generally fairly small except for early and late in the 2010/11
apparition when we were attempting to extend the temporal coverage as
much as possible and Hartley 2 was quite faint.

\section{NARROWBAND PHOTOMETRY}
\label{sec:photometry}
\subsection{The Photometry Data Set}
We first present our photometric results for the three apparitions 
for which we have data. In Table~\ref{t:phot_flux} we list emission band and continuum 
fluxes (as logarithms) along with the resulting gas column abundances, 
M($\rho$), while the derived gas production rates and \afrho\ 
values for each continuum point are given in Table~\ref{t:phot_rates} as well as the 
vectorial-equivalent water production rate in the rightmost column. 
Note that although the 1-$\sigma$ uncertainties are unbalanced in 
log-space, we only list the ``+'' log value for clarity and to save space; 
the ``$-$'' values can be computed knowing that the percent uncertainties 
are balanced.



\subsection{Temporal Behavior}

\begin{figure}
  \centering
  \includegraphics[width=170mm]{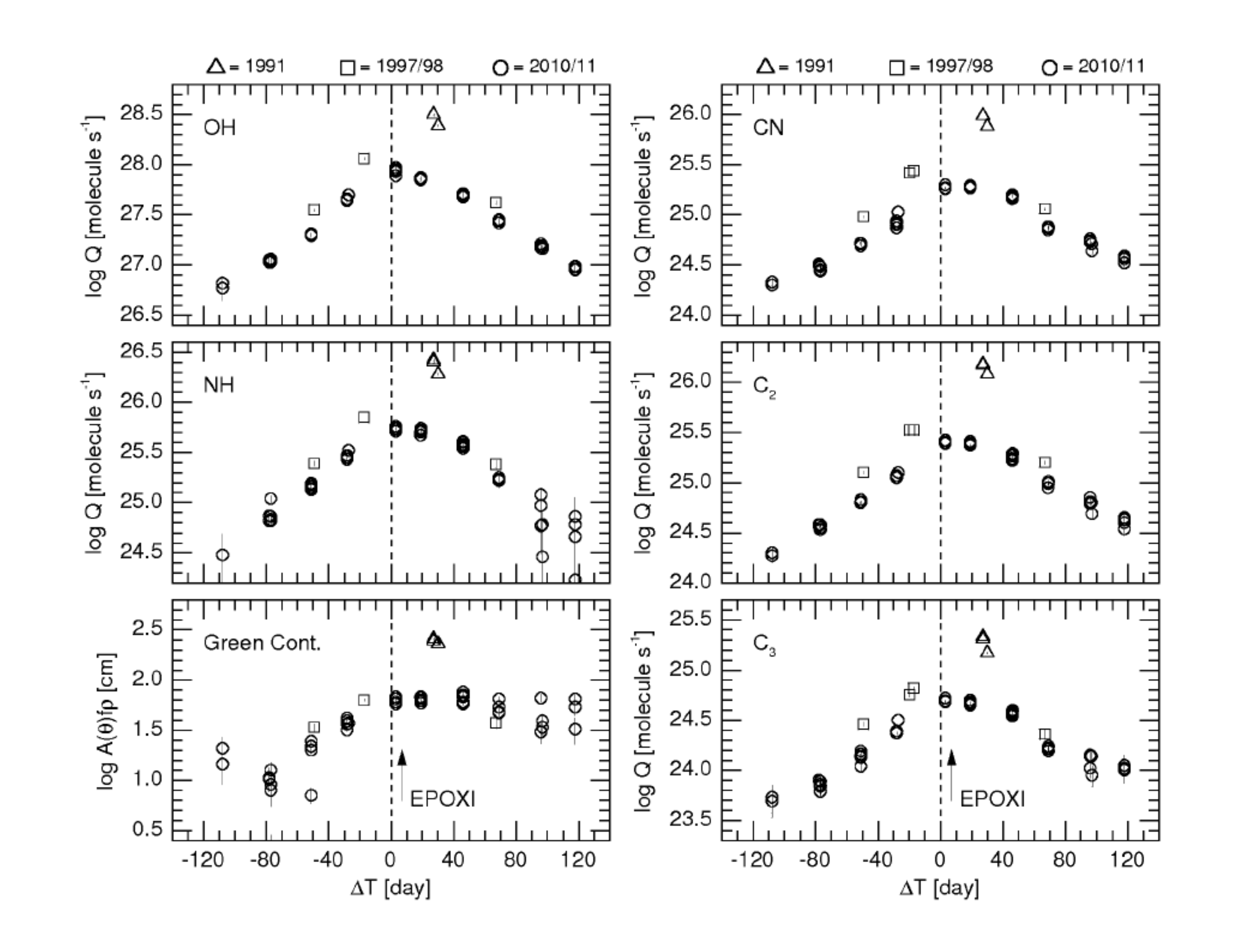}   
  \caption[Gas and dust production rates]{Log of the production rates for each observed molecular species
and \afrho\ for the green continuum plotted as a function of time from 
perihelion. Data points from the 1991 apparition are shown as triangles, 
those from 1997/98 are given as squares, and the recent 2010/11 data are 
shown as circles. Error bars are plotted; in cases where they are not 
visible it is because they are smaller than the symbols. The time of the 
EPOXI spacecraft flyby is shown with an 
arrow. Note the large asymmetries around perihelion for all 
species, and the much shallower dust slope as compared to the gas species 
following perihelion. Although the perihelion distance increased at 
each successive apparition, the drop in production rates with apparition 
is much larger than can be explained by the relatively small drop in 
solar illumination.}
  \label{fig:phot1}
\end{figure}

The derived production rates for each gas species and \afrho\ for the green 
continuum are plotted as logarithms in Figure~\ref{fig:phot1}
as a function of time from perihelion. The two most obvious characteristics 
are that the earlier apparitions generally have higher production 
rates than the most recent apparition and that all species exhibit 
a bulk asymmetry about perihelion, with values higher after 
but by varying amounts. As is usual for Jupiter-family comets 
(cf. \citealt{ahearn95}), the heliocentric distance dependancies 
for the gas species (log $Q$ vs log \rh ) are significantly steeper than 
a canonical \rh$^{-2}$ (or \rh$^{-2.5}$ if adjusting for gas outflow velocities). 
Based solely on the 2010/11 apparition, the slopes for the 
carbon-bearing species range from $-$3.4 to $-$4.1 before perihelion and 
from $-$3.2 to $-$3.5 after, while OH is $-$4.6 prior to and $-$4.0 after, 
and NH has an uncertain value of $-$7.1 before and $-$4.9 after perihelion 
(see Table~\ref{t:table5}). 



The \rh -dependancy for the dust, $Af\rho$, before perihelion is quite similar 
to those exhibited by the carbon-bearing species, especially if 
one normalizes to 0\deg\ phase angle. Specifically, the three continuum 
points have values for the slopes from $-$2.8 to $-$3.6 (self-consistent 
given the large uncertainties) when using \afrho\ while somewhat 
higher values from $-$3.4 to $-$4.1 using $A$(0\deg)$f$$\rho$.
Although the adjustment for phase angle has a larger effect after 
perihelion -- going from $-$1.0 to $-$1.2 without adjustment to $-$1.7 to $-$1.9 
when normalized -- as seen in other comets, including 19P/Borrelly \citep{schleicher03b}
and 67P/Churyumov-Gerasimenko \citep{schleicher06b}, the \rh -dependancies 
for dust are much shallower than that of the gas species following 
perihelion. As we suggested in Borrelly's case, the very 
shallow slope after perihelion is likely due to the release of larger, 
very slow moving grains near perihelion and peak water production, and 
these grains remain in the inner coma far longer than the more typical 
micron-sized dust particles. 
Note that $Af\rho$ also exhibits the usual trend with aperture size, 
where larger apertures yield smaller values, indicating that the 
dust spatial distribution with projected distance from the nucleus is 
steeper than the canonical 1/$\rho$. Since our projected aperture sizes 
were generally larger when the comet was at larger heliocentric 
distances, the \rh-dependences just-discussed would have been even 
shallower had an adjustment for aperture size been performed, 
thereby magnifying the post-perihelion effect. 

Returning to the earlier apparitions of the 1990s, we first note that 
all observational sets from 1997/98 are higher than corresponding data 
from 2010 except for the sole continuum point after perihelion that 
can be explained by a combination of aperture and phase angle effects. 
The measured offset between these two apparitions ranges between about a 
factor of 1.5 and 2.0 among all species, far more than can be explained 
by the change in solar flux caused by the increase in perihelion distance 
from 1.032 AU in 1997 to 1.059 AU in 2010. While our data are even
sparser in 1991, with only three data sets on two nights, the measured 
offset is much greater, corresponding to a factor of 3--4 as compared to 2010. 
Again, this is far greater than expected due to solar illumination -- 
perihelion was 0.953 AU in 1991, resulting in only a 25\% difference. 
To examine this further, we next intercompare a variety of water 
measurements from all three apparitions.

\subsection{Water Production}

\begin{figure}
  \centering
  \includegraphics[width=160mm]{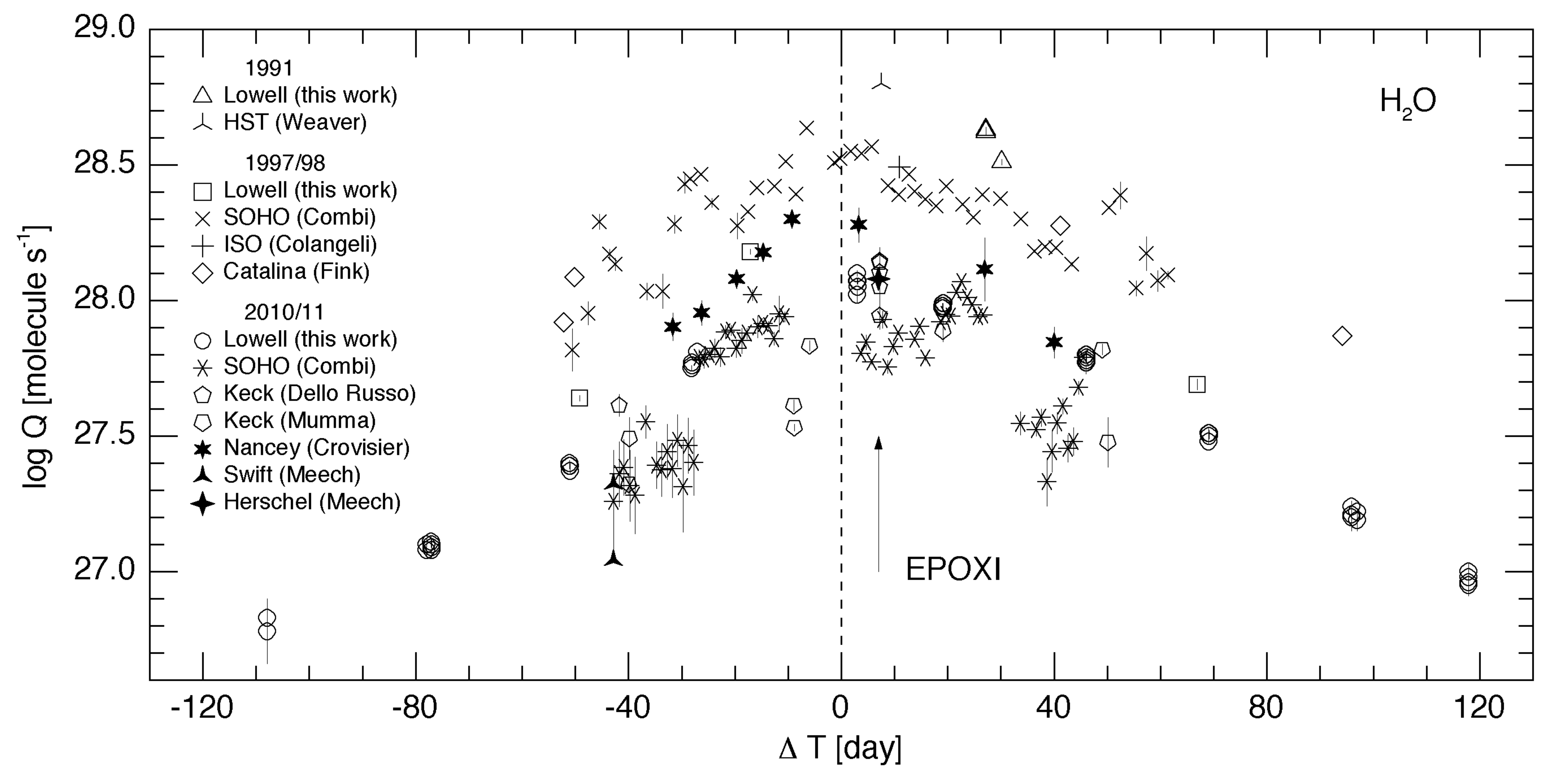}   
  \caption[Log of water production rate relative to perihelion]{Log of the production rates for water plotted as a function of
time from perihelion. Our values, based on OH, are shown with the same 
symbols as Figure~\ref{fig:phot1}. Other OH-based results include {\it HST}/FOS data by 
\citet{weaver94}, data from the {\it Swift} satellite (Bodewits private communication; \citealt{meech11}), 
and radio OH measurements from Nancey \citet{crovisier12}.
Forbidden oxygen data were obtained by \citet{fink09} at the 1997 apparition, 
while the granddaughter hydrogen was measured using the Lyman alpha 
line with {\it SOHO}/SWAN (\citealt{combi11a}; \citealt{combi11b}). In the IR, water 
measurements have become more common, with space-based data from 
{\it ISO}/ISOPHOT \citep{colangeli99} and {\it Herschel} \citep{meech11}, 
and ground-based data with Keck/NIRSPEC \citep{iauc9171,dellorusso11,mumma11}. See the key to associate the symbols with these
references; only the first author is listed in the key due to space 
constraints. As discussed in the text, the ensemble of data indicate 
a larger drop in production rates between 1997/98 and 2010/11 and a larger 
amount of variability than we inferred from our own data alone. 
}
  \label{fig:phot2}
\end{figure}

As indicated in Section~\ref{sec:phot_obs}, we can compute vectorial-equivalent 
water production rates from our Haser-model OH production rates, and 
we plot these in Figure~\ref{fig:phot2}. We also overlay all other published water 
production values, including when available the uncertainties. 
We were surprised at the large amount of dispersion, even when only 
examining one apparition at a time. While some of this dispersion is 
likely due to differences in modeling parameters, some due to 
species-related issues (data are from direct water measurements in 
the IR, OH in the near-UV and radio, and H in the far-UV), and some due 
to varying dilutions by older material associated with an extremely large 
range of effective aperture sizes for the measurements, there 
remains a great deal of variability within some individual data sets, 
especially as compared to the apparent smooth curve with time evident 
in our own data.  


Looking in detail first at the 2010/11 apparition, the variability 
seen in the most comprehensive data set from {\it SOHO}/SWAN \citep{combi11b} 
is sometimes very small ($\Delta$T = $-$27 to $-$10 day) but other times 
more than a factor of two in only a few days. 
And while it first appears that our data are systematically higher 
than that from {\it SOHO}/SWAN, a closer examination reveals excellent 
agreement when data were obtained near-concurrently, such as at $\Delta$T
of $-$28/$-$27 day, +19 day, and +46 day (where the {\it SOHO}/SWAN point is 
nearly invisible due to the overlap). Other data sets, such as Keck 
\citep{mumma11}, also suggest considerable short-term variability 
during some intervals in the apparition. Based on the ensemble of data, we 
conclude that our own smooth curve during 2010/11 was partly caused by having 
only one successful night of photometry on the majority of our observing 
runs, due to a combination of relatively poor weather and our focus towards 
obtaining imaging data (Section 4).  Another factor is simple random chance; 
for instance our data sets from Nov 16 (+19 days) and Dec 13 (+46 days) are in 
complete agreement with other data on those nights. However, in neither case 
are these data similar in value to data obtained during the prior $\sim$10 days.

Turning next to 1997/98, {\it SOHO}/SWAN \citep{combi11a} again provides 
the most complete record, with one interval after perihelion showing 
a long steady decline while at other times much larger variability is evident. 
Also, in 1997, {\it SOHO}/SWAN measurements are systematically higher than 
ours, opposite of the case in 2010, but our two pre-perihelion data 
points each occur at minima of the {\it SOHO}/SWAN data set. 
When looked at together, it becomes clear that there was a much 
greater drop in production rates between 1997 and 2010, by about a 
factor of 3, than our data alone suggested. 
By extension, this would
also indicate a smaller change took place between 1991 and 1997, and this 
is consistent with the only other published data point from 1991 
({\it HST}; \citealt{weaver94}), along with an unpublished {\it IUE} measurement 
taken within the week that was stated by Weaver et al. to be similar in value. 
Allowing for the change in perihelion distances by comparing the 1991 data 
with an extrapolated heliocentric distance trend post-perihelion, our 1991 data 
are about 50\% higher while the {\it HST} (and {\it IUE}) points are nearly 90\% higher 
than 1997. While the sparseness of the 1991 data and the large variability 
that is expected to be present make this result somewhat uncertain, the 
data are suggestive that water production rates were $\sim$70\% greater 
in 1991 than in 1997.

From these water data, we conclude that there was a large ($\sim$3$\times$) decrease
in water vaporization between 1997 and 2010 (an interval of two orbits), 
and a smaller ($\sim$1.7$\times$) decrease between 1991 and 1997 (a single orbit), 
which implies a relatively consistent drop of about 40\% from one apparition 
to the next. There is also considerable evidence for a stochastic 
amount of variability in water production as a function of time. 
There were intervals when variability was 
small, smooth, and exhibited trends consistent with the change in distance from 
the Sun, and other times when variability was far larger and 
apparently sporadic. Another striking comparison is the opposite trends 
evident in the SOHO data sets in the 3-4 weeks immediately after perihelion 
in 1997 versus 2010. Given the evidence from the \epoxi\ mission that 
there were several isolated source regions on the nucleus, coupled 
with evidence for a complex rotational state and a changing rotational 
period of at least one of the rotational components, we conclude 
that at least one source region experiences a very complicated 
solar illumination from a combination of these rotational effects. We further 
suggest that the intervals of smooth trends interrupted by intervals of 
much larger variability may be directly associated with the beat between 
the two periods, and how this beat changed with the changing period(s). 

The absolute values of the water production can also yield a nominal 
surface area of vaporization. Our values in 2010 near perihelion and the 
\epoxi\ encounter imply an active area of 3--4 km$^{2}$, just below the 
measured nucleus surface area of $\sim$4.2 km$^{2}$ (based on a mean radius 
of 0.58$\pm$0.02 km from \citealt{ahearn11a}). Note that far more surface 
area than exists on the nucleus would have been required to produce 
the amount of water detected in the 1990s. The significance of this 
will be returned to in Section~\ref{sec:disc}.

\subsection{Composition}
\label{sec:composition}
The composition of Hartley 2, based on the abundance ratios of the 
minor gas species computed using the ratios of their respective 
production rates with respect to OH, are given in Table 5. 
Here we also provide our standard uncertainties, the $\sigma$ of the data, 
which we have tabulated in our previous single comet papers, 
along with the $\sigma$ of the mean. The former describes the dispersion 
of the data, due to observational errors, modeling effects, and 
intrinsic variations in the comet itself, while the latter better 
describes how well each mean ratio is determined. 

Hartley 2's composition clearly places it in the middle of the 
``typical'' classification from \citet{ahearn95}, consistent 
with our result from the 1991 apparition that was included in A'Hearn et al. 
Using this same classification scheme, \citet{lara11} find 
the same result from their spectroscopic observations. 
The comet is also classified as ``typical'' by \citet{fink09} using his 
own definition of compositional classes. 

Determining a single mean value for the dust-to-gas ratio is more 
problematic for several reasons, including the trend in $Af\rho$ values with 
aperture size, the much shallower \rh -dependence following perihelion, 
and phase angle effects. Since the first two issues are incorporated 
in our calculation of the uncertainties of the data for the dust-to-gas 
ratio, we make no further adjustments. Regarding phase angle effects, we 
compute both an unadjusted value, i.e. \afrho/$Q$(OH), for direct 
comparison to other comets in \citet{ahearn95}, along with a 
normalized value to 0\deg\ phase angle. Our results for the green 
continuum were $-$25.84 and $-$25.44, respectively, corresponding to 
a relatively low dust-to-gas ratio, especially as compared to other 
Jupiter-family comets. 
Finally, based on the three continuum points we measure, the dust 
is nearly grey in color, with reddening of less than 10\% per 1000~\AA\ 
through most of the apparition. However, at the end of October, when 
our projected apertures were smallest, the color in the UV was as high
as 30\% per 1000~\AA. This result is consistent with that by
\citet{lara11} who detect a stronger reddening in the inner-most coma, 
but note that the reddening decreases with distance and also varies 
with direction in the coma and that the dust is blue in some locations.

\section{COMA MORPHOLOGY}
\label{sec:gas_morphology}
\subsection{Review of CN Morphology}
Paper 1 focused on Hartley 2's CN coma morphology. We summarize the relevant results here to place the morphology observed in OH, NH, C$_3$, C$_2$, and dust into context. We observed two CN features, generally towards the north and south. In August and September only the northern feature was seen, and had the appearance of a nearly face-on spiral. In October, November, and December the morphology was roughly an hourglass shape, with outward motion visible and the relative brightnesses of the northern and southern features varying.
In January the signal-to-noise was poor and only the southern feature was detected. The CN morphology varied smoothly during a night and similar, but not identical, morphology was seen on subsequent nights. The similar morphology allowed us to estimate a rotation period during the August to November runs, and the period increased by $\sim$2 hr (from $\sim$16.7 hr to $\sim$18.7 hr) during this time. The differences in morphology implied the existence of small deviations from a principal axis rotation. The morphology repeated much better 3, 6, 9... cycles apart than it did 1, 2, 4, 5... cycles apart, implying a component rotation period $\sim$3$\times$ the ``primary'' rotation period. Similar morphology was reported by \citet{samarasinha11}, \citet{lara11}, and \citet{waniak12}, and the changing rotation period and non-principal axis rotation were confirmed by other observers (cf. \citealt{cbet2589,ahearn11a,drahus11,samarasinha11}).

\subsection{Coma Morphology of C$_3$, C$_2$, OH, and NH}
\begin{figure}
  \centering
  \includegraphics[width=120mm]{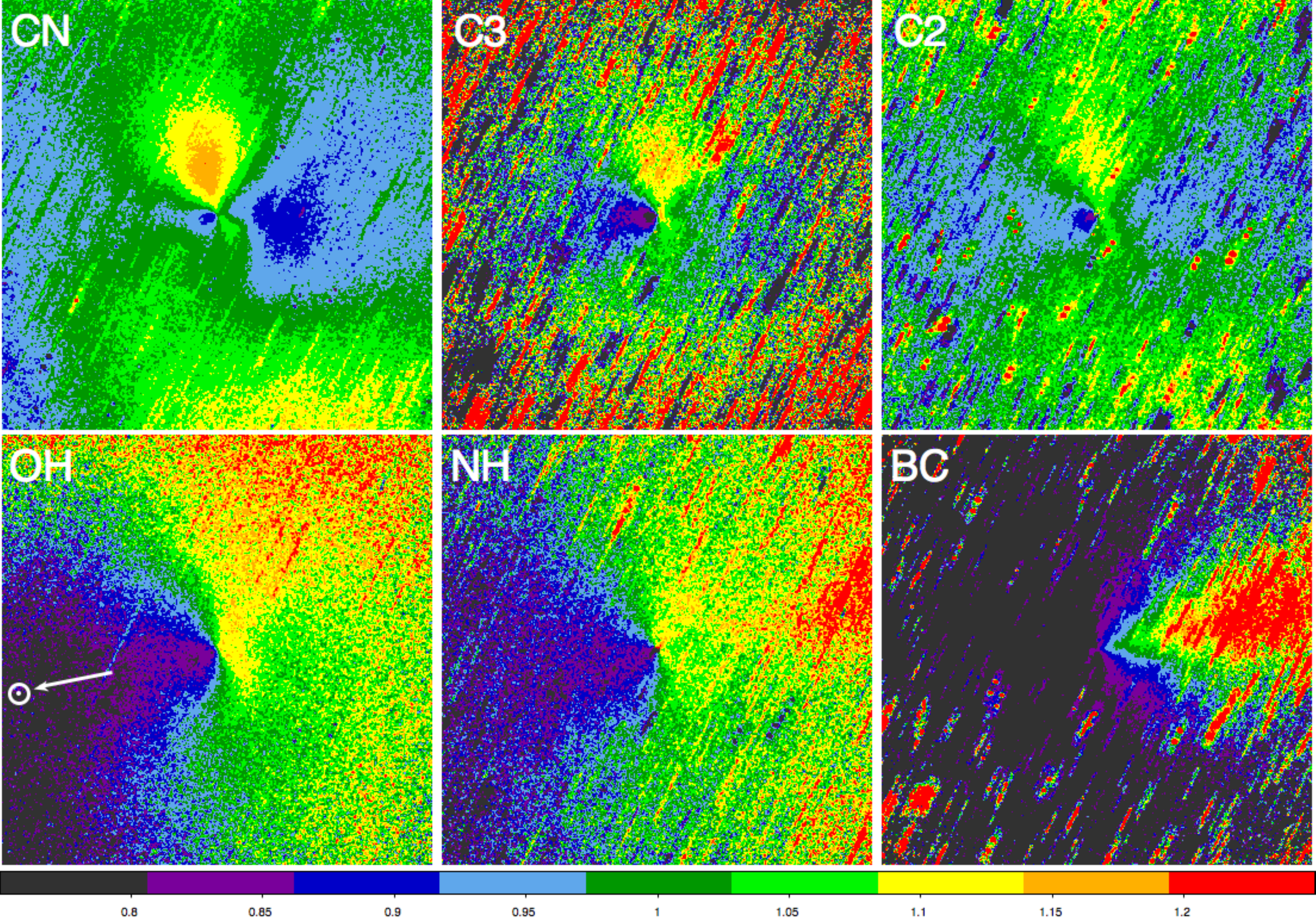}  
  \includegraphics[width=120mm]{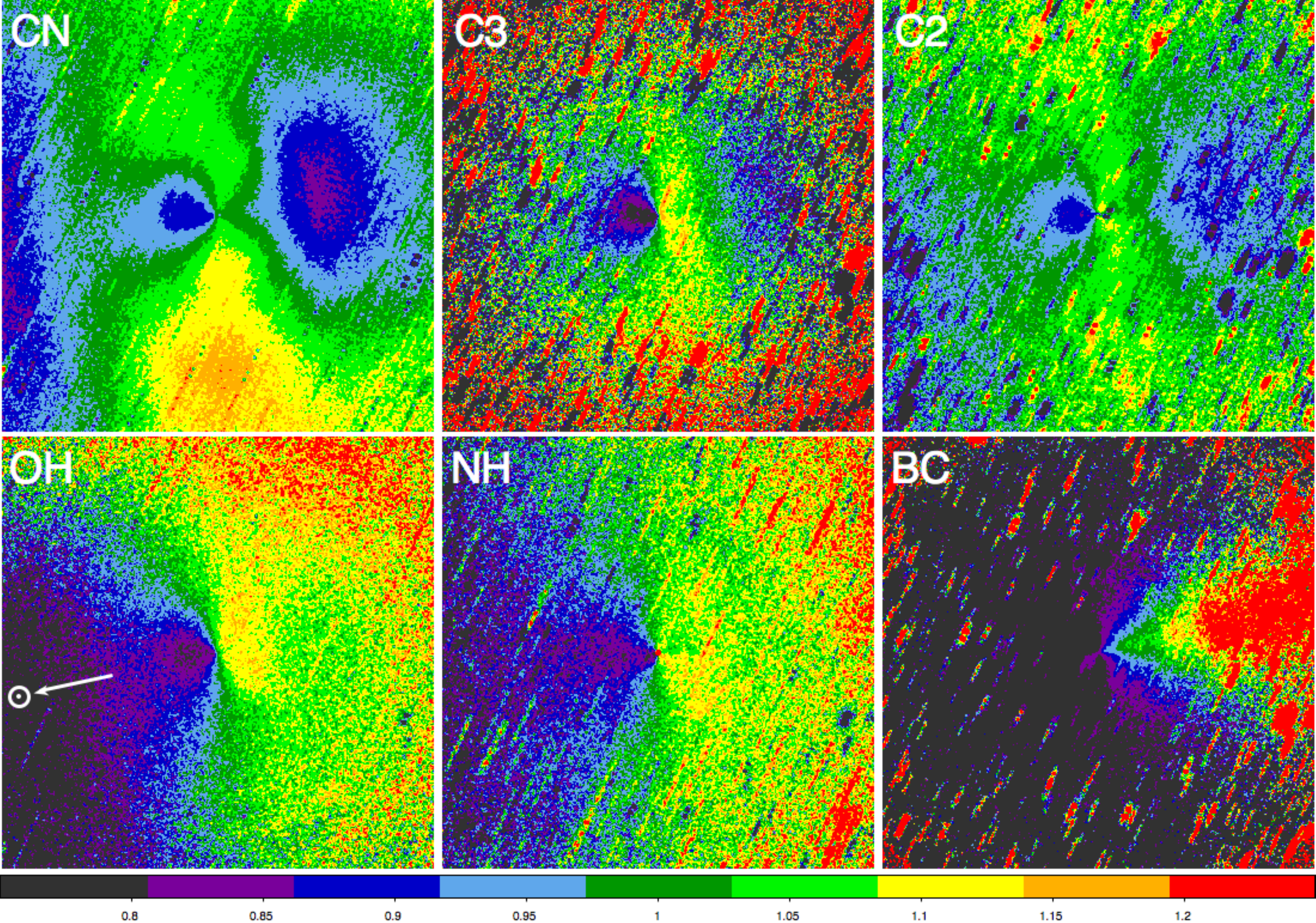}  
  \caption[Gas and dust coma morphology]{Gas and dust coma morphology of 103P/Hartley 2 on 2010 November 2 (top panel) and 2010 November 3 (bottom panel). The bandpass is given in the top left of each image with BC denoting blue continuum (e.g., dust). Each image is a decontaminated pure gas or dust image centered on the nucleus and enhanced by division of an azimuthal median profile then smoothed with a boxcar smooth. Each image is approximately 64,000 km across with north up and east to the left. The direction to the Sun is indicated in the OH frames.
The color scale is given at the bottom; the stretch is the same for all gas images, and a different stretch is used for both BC images. Trailed stars are visible as streaks extending from the northwest to the southeast. Note that some artifacts of the enhancement are visible as ring-like structures}
  \label{fig:gas_morph1}
\end{figure}

CN is generally the gas species of choice for investigations of gas coma features with narrowband filters because it is vastly brighter than OH or NH and has a better contrast relative to the dust than C$_2$, C$_3$, or NH. CN jets were first detected in the coma of 1P/Halley \citep{ahearn86} and jets and/or fans have now been detected in numerous other comets, e.g., C/1995 O1 Hale-Bopp \citep{woodney02}, C/2004 Q2 Machholz \citep{farnham07a}, 8P/Tuttle \citep{waniak09}, and C/2007 N3 Lulin \citep{knight09a}. The coma morphology of the other gas species has only been published for a few very bright comets, e.g., 1P/Halley \citep{cosmovici88,schulz95} and C/1995 O1 Hale-Bopp \citep{lederer97}. Hartley 2's very small geocentric distance during the 2010/11 apparition allowed us to observe it with unusually high spatial resolution, making a multiwavelength study of coma morphology highly desirable. As listed in Table~\ref{t:imaging_circ}, we obtained C$_2$ and C$_3$ images on photometric nights from August through November, OH images on photometric nights from September through November, and NH images on photometric nights in October and November (although the lone October NH image is unusable due to a tracking error). The comet was easily detectable in all of these raw images and after removal of dust contamination. However, sufficient signal-to-noise to allow meaningful analysis of the enhanced images was only possible in October and November for species other than CN.
We show CN, C$_3$, C$_2$, OH, NH, and dust on November 2 and 3 in Figure~\ref{fig:gas_morph1}; comparable data were obtained on October 16 and November 7 but are not shown.


As shown in Figure~\ref{fig:gas_morph1}, the C$_3$ and C$_2$ morphology looked generally similar to the CN morphology. On November 2, all three species exhibited an hourglass-like shape, with a stronger feature to the north and a fainter feature to the south; the southern feature can be seen leaving the nucleus towards the southwest. On November 3, the CN, C$_3$, and C$_2$ showed a somewhat different hourglass shape, with the the southern feature much brighter relative to the northern feature than on November 2. All three species were much fainter along the sun-tail line (roughly east-west) than along the axis of the hourglass feature. The CN signal was typically lowest in the tailward direction, whereas the C$_3$ and C$_2$ signals were typically slightly fainter in the sunward direction. The overall bulk brightness of C$_3$ and C$_2$ varied in correlation with the bulk CN brightness, e.g., generally brighter to the north on November 2 and generally brighter to the south on November 3. 

\begin{figure}
  \centering
  \epsscale{1.0}
  \includegraphics[height=170mm,angle=270]{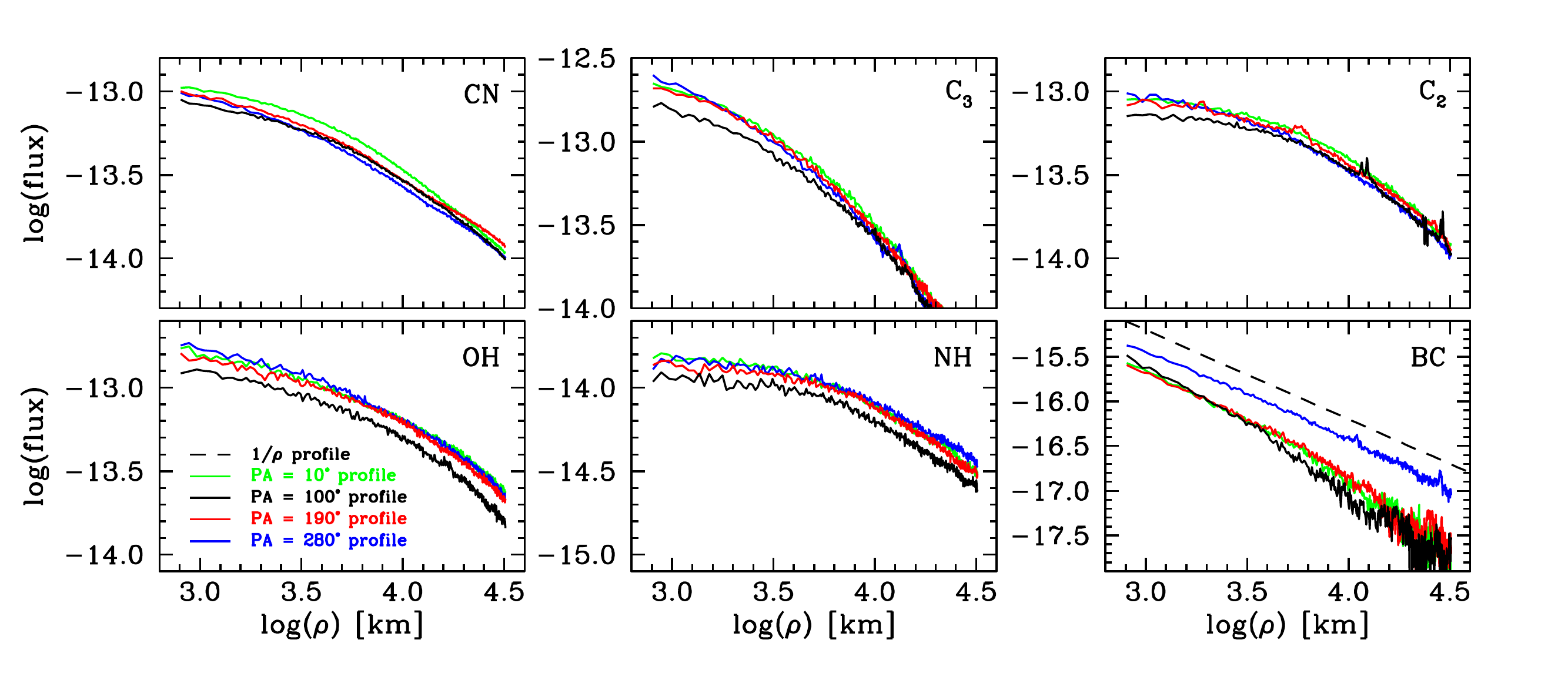}   
  \caption[Gas and dust profiles]{Log of the gas and dust flux on 2010 November 2 as a function of the log of the distance from the nucleus in the sunward direction (black), anti-sunward direction (blue), and orthogonal to the sunward line to the north (green) and to the south (red). Each profile is the mean of all points a distance $\rho$ from the nucleus within a 10$^\circ$ wide wedge centered on a given position angle (PA). The wedges are centered at PAs 10$^\circ$ (green), 100$^\circ$ (black), 190$^\circ$ (red), and 280$^\circ$ (blue). The sun is at a PA of 102$^\circ$. The species is given in the top right corner of each plot. The flux has units of erg cm$^{-2}$ s$^{-1}$ for the gas species and erg cm$^{-2}$ s$^{-1}$ \AA$^{-1}$ for BC. A 1/$\rho$ profile is shown as a dashed line in the BC plot. Note that the slopes of the gas species can be compared directly, as they are plotted with the same $\Delta$log(flux), but with different ranges.}
  \label{fig:profiles}
\end{figure}

In order to quantify the differences between the gas species, we show their spatial profiles in Figure~\ref{fig:profiles}. We plot the flux (as measured on decontaminated, but not enhanced, images) on November 2 as a function of distance from the nucleus in 10$^\circ$ wide wedges centered along position angles (PAs) of: 10$^\circ$ (perpendicular to the sunward direction, to the north), 100$^\circ$ (the sunward direction), 190$^\circ$ (perpendicular to the sunward direction, to the south), and 280$^\circ$ (the tailward direction). These profiles extend to the edges of the images shown in Figure~\ref{fig:gas_morph1}. The inner $\sim$800 km are not plotted as there are very few pixels in each wedge at these distances, causing small fluctuations to be exaggerated. The northern and southern CN profiles were $\sim$20\% and $\sim$10\% brighter than the sunward profile, respectively. The CN tailward profile was roughly the same brightness as the sunward profile. The C$_3$ and C$_2$ profiles were both brightest to the north, followed by the south, then tailward, and finally sunward. The north, south, and tailward profiles were 10--30\% brighter than the sunward profile. Note that these brightnesses are along the line of sight and are products of projection effects, each species' parentage, and excess velocities acquired. Thus, brightness enhancements in a particular direction relative to the sunward direction (which did not have any obvious jets) are likely only lower limits, and the actual ratio of the material originating from the nucleus in each direction may be substantially higher.


Dividing the C$_2$ and C$_3$ images by the CN images revealed minimal differences in their morphologies, with the primary feature (the hourglass shape) disappearing in quotient images. As is clearly evident in Figure~\ref{fig:profiles}, the C$_3$ does not extend as far as the other gas species.
This was partially due to the higher signal-to-noise in CN and C$_2$ as compared to C$_3$, however, it was mostly due to the differing lifetimes of the species and their respective parentages; C$_3$ and its parents have much shorter lifetimes than either C$_2$ or CN and their respective parents (cf. \citealt{ahearn95}). C$_2$ appeared quite similar to CN but was more diffuse because it has multiple parents and grandparents, whereas CN is primarily a daughter species. The multiple parentages of C$_2$ cause its radial profile to be flatter than the profiles of either CN or C$_3$.

Owing to the shorter observing window when the comet was brightest and the desire to acquire images in a large number of filters, we usually only acquired one or two C$_2$ or C$_3$ images per night. Thus, we cannot investigate how the morphology of these species compares as a function of rotation period. However, on every night in which we saw CN, C$_3$, and C$_2$ with sufficient signal-to-noise to discern varitions in the coma, the morphologies appeared to be completely consistent except for the parentage and lifetime effects discussed above. This leads us to conclude that the C$_3$ and C$_2$ species exhibit the same rotational and seasonal morphology as CN and therefore originate from the same source region(s) on the surface.

Near the nucleus, the NH displayed some of the hourglass morphology of CN, C$_3$, and C$_2$. The features appeared to originate from separate northern and southern sources, although the distinction between the features in the tailward direction was less obvious than for CN, C$_3$, or C$_2$. The relative brightness of the northern and southern features varied in correlation with changes in the brightness of these features in CN. NH had a strong asymmetry in brightness of the sunward and tailward hemispheres, with the tailward hemisphere always being substantially brighter. The bulk brightness of the NH images displayed little variation from night to night. The radial profiles demonstrate the extent of the tailward brightness enhancement, as the north, west, and south profiles all remained at least 20\% higher than the eastern (sunward) profile to the edge of the images displayed in Figure~\ref{fig:gas_morph1}. As with C$_2$, the NH radial profiles were rather flat due to its being primarily a granddaughter species. We used the photometry to determine the amount of reddening to apply in order to properly remove the underlying continuum from the NH. However, it is likely that the continuum was somewhat over-removed in the innermost coma ($<$3000 km) because the dust was increasingly reddened at progressively smaller distances from the nucleus, resulting in the unusually flat profile.

The morphology most different from CN was OH. The OH morphology near the nucleus showed only hints of the hourglass shape. Unlike the other gas species, the feature did not have distinct northern and southern components, but was essentially continuous from the north through the west (tailward) to the south. The relative brightnesses of the north and south features of the hourglass vary much less than the other gas species. Like NH, the tailward hemisphere was much brighter than the sunward hemisphere. Furthermore, at larger distances the OH was brighter toward the north on all four nights in November, and the bulk brightness did not vary significantly from night to night. The tailward profile was $\sim$35\% higher than the sunward profile, while the northern and southern profiles were 20--25\% higher than the sunward profile.

\begin{figure}
  \centering
  \epsscale{0.5}
  \includegraphics[width=80mm]{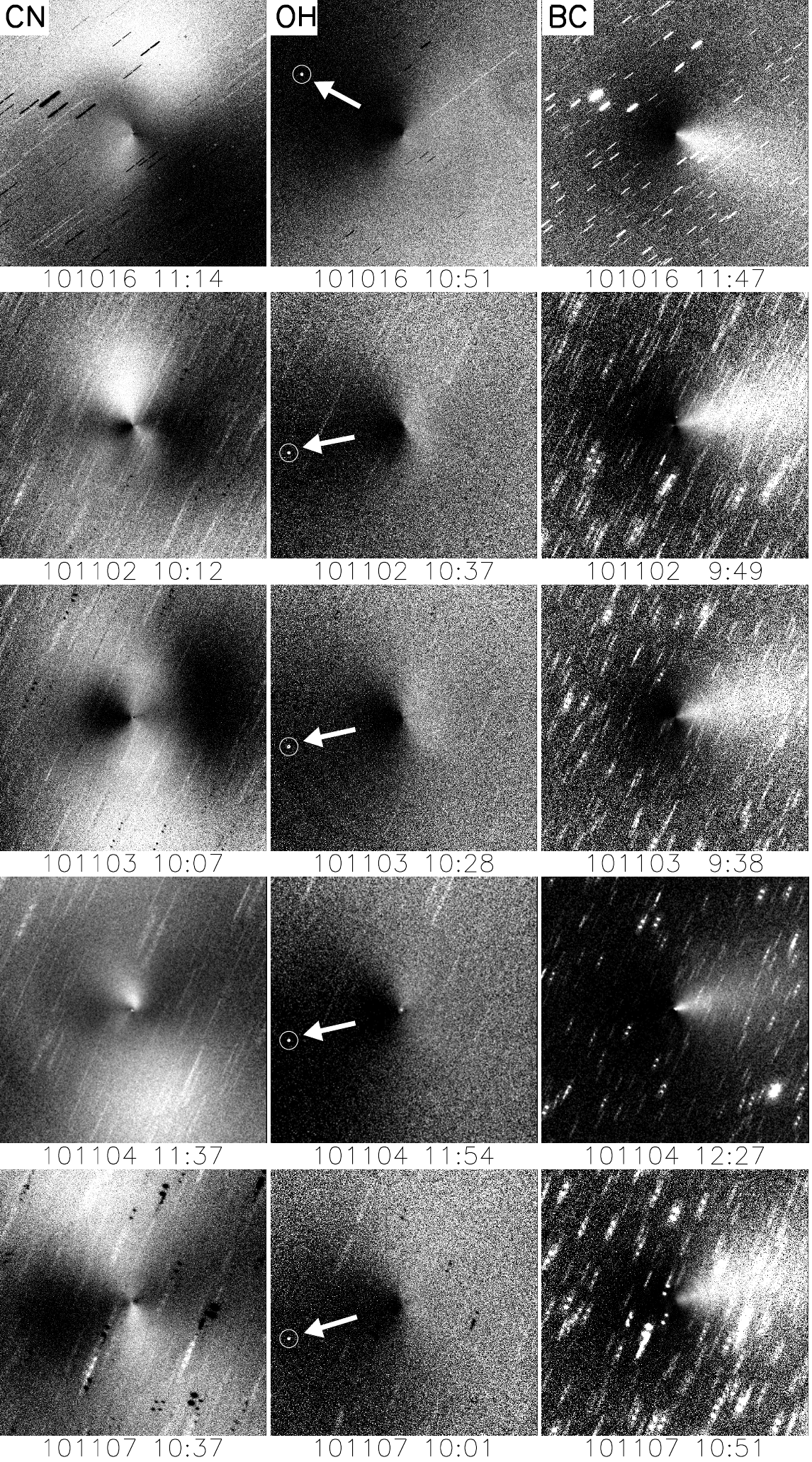}   
  \caption[Gas and dust coma morphology]{CN, OH, and dust continuum of 103P/Hartley 2 in 2010 October and November. The left column is CN, the middle column is OH, and the right column is dust (blue continuum; BC). The date (YYMMDD) and UT time (HH:MM) of the midpoint of each image is below the image. Each image is a decontaminated pure gas or dust image except for 2010 November 4 (the night of the \epoxi\ flyby) when it was not photometric, and the original image is shown after bias removal and flat-fielding. Each image is centered on the nucleus and enhanced by division of an azimuthal median profile, is approximately 50,000 km across, and has north up and east to the left. The direction to the Sun is indicated on the OH image for each day. All of the images in a given filter have the same stretch, but the stretch varies from filter to filter. In all cases white is bright and black is dark. Stars appear as trailed streaks, and some images have faint circular artifacts from the enhancement process. The OH morphology is distinctly different from the CN morphology and is concentrated in the anti-sunward hemisphere, suggesting that a substantial amount of OH is coming from small, icy grains.}
  \label{fig:gas_morph2}
\end{figure}

The striking differences between the coma morphologies of OH and CN was seen on every night in which we have usable OH data, as shown in Figure~\ref{fig:gas_morph2}. In each of these images the OH signal near the nucleus showed hints of the hourglass shape seen in CN (although the faint CN feature near the nucleus and towards the southeast on November 4 is absent in OH). The tailward OH hemisphere was always much brighter than the sunward hemisphere, and the dust tail (BC) almost perfectly bisected the OH distribution in all images, including October 16 when the geometry was substantially different. The hemispheric brightness asymmetry and the lack of variation in the bulk brightness from night to night suggest that there is a smearing across rotational phase in the OH signal, and that the changing viewing geometry was primarily responsible for the changes in OH morphology between October and November.


As with C$_2$ and C$_3$, we typically only acquired one or two sets of OH and NH on a given night so we cannot determine the morphology throughout an entire rotation cycle. However, during each night in which we observed OH and NH, their bulk coma morphologies looked similar -- concentrated in the tailward hemisphere -- and distinctly different than CN, C$_2$, or C$_3$. The consequences of these determinations will be returned to in Section~\ref{sec:disc}.

\subsection{Dust Morphology}
\label{sec:dust_morphology}
\begin{figure}
  \centering
  \epsscale{1.0}
  \includegraphics[width=170mm]{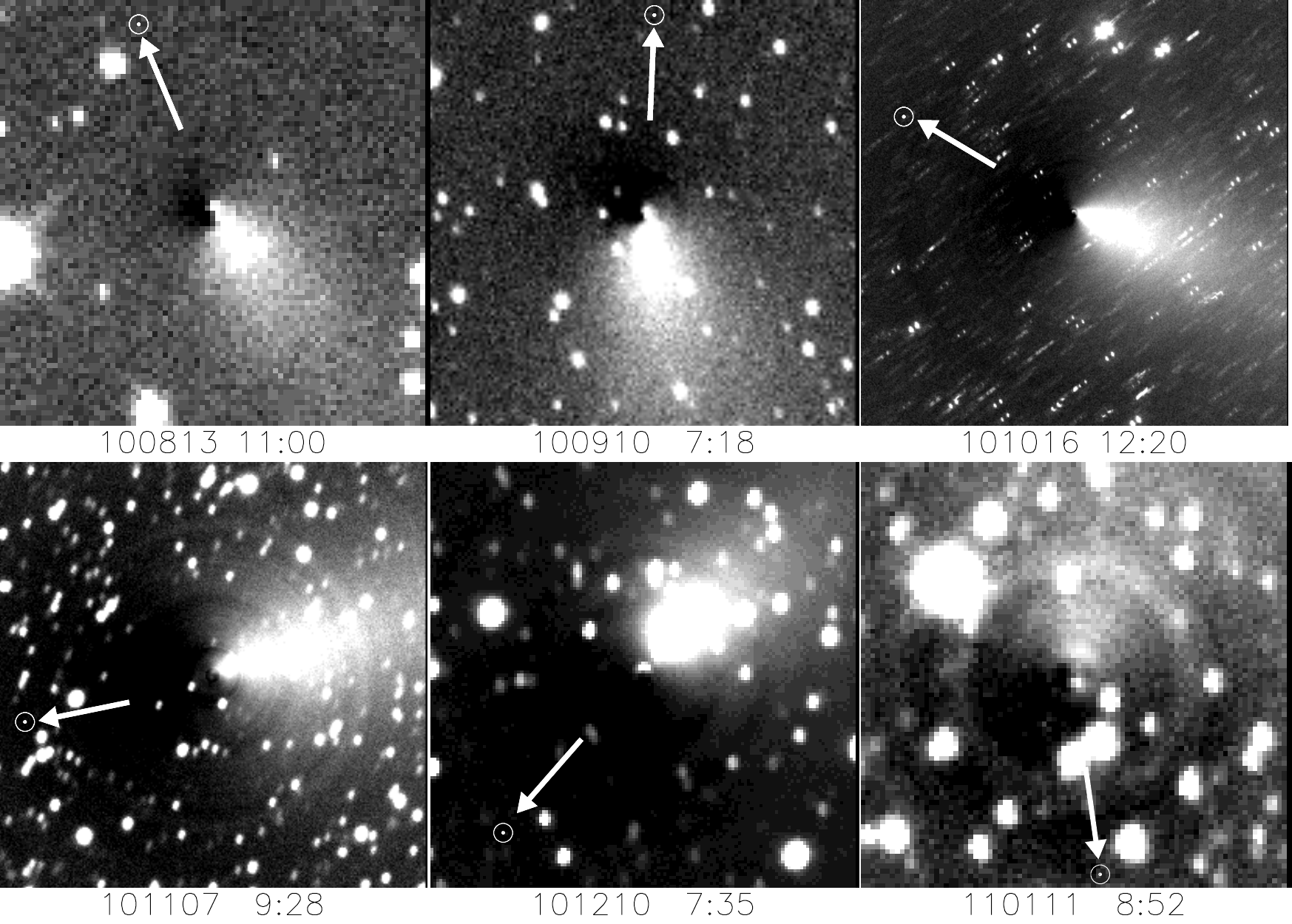}   
  \caption[R-band morphology]{Evolution of R-band (dust) morphology of 103P/Hartley 2 from 2010 August until 2011 January. All images are R-band. All other details are as given in Figure~\ref{fig:gas_morph2}. Note that some artifacts of the enhancement are visible as ring-like structures.
}
  \label{fig:r_morph}
\end{figure}

The dust morphology of Hartley 2 was dominated by the tail throughout the apparition. Figure~\ref{fig:r_morph} shows enhanced R-band images monthly from 2010 August through 2011 January. This same shape was seen, albeit with lower signal-to-noise,  in the narrowband continuum filters: UV, blue, green, and red. In all cases the dominant feature was roughly straight and in the anti-sunward direction (the PA of the Sun is labeled on each panel). It does not vary with rotational phase and is consistent with the expected position of the dust tail, confirming that it is the dust tail. Overall, the brightness of the dust systematically decreased as a function of PA from the tailward direction to the sunward direction. Radial profiles of the four fundamental directions are shown in Figure~\ref{fig:profiles}.

\begin{figure}
  \centering
  \epsscale{1.0}
  \includegraphics[width=60mm]{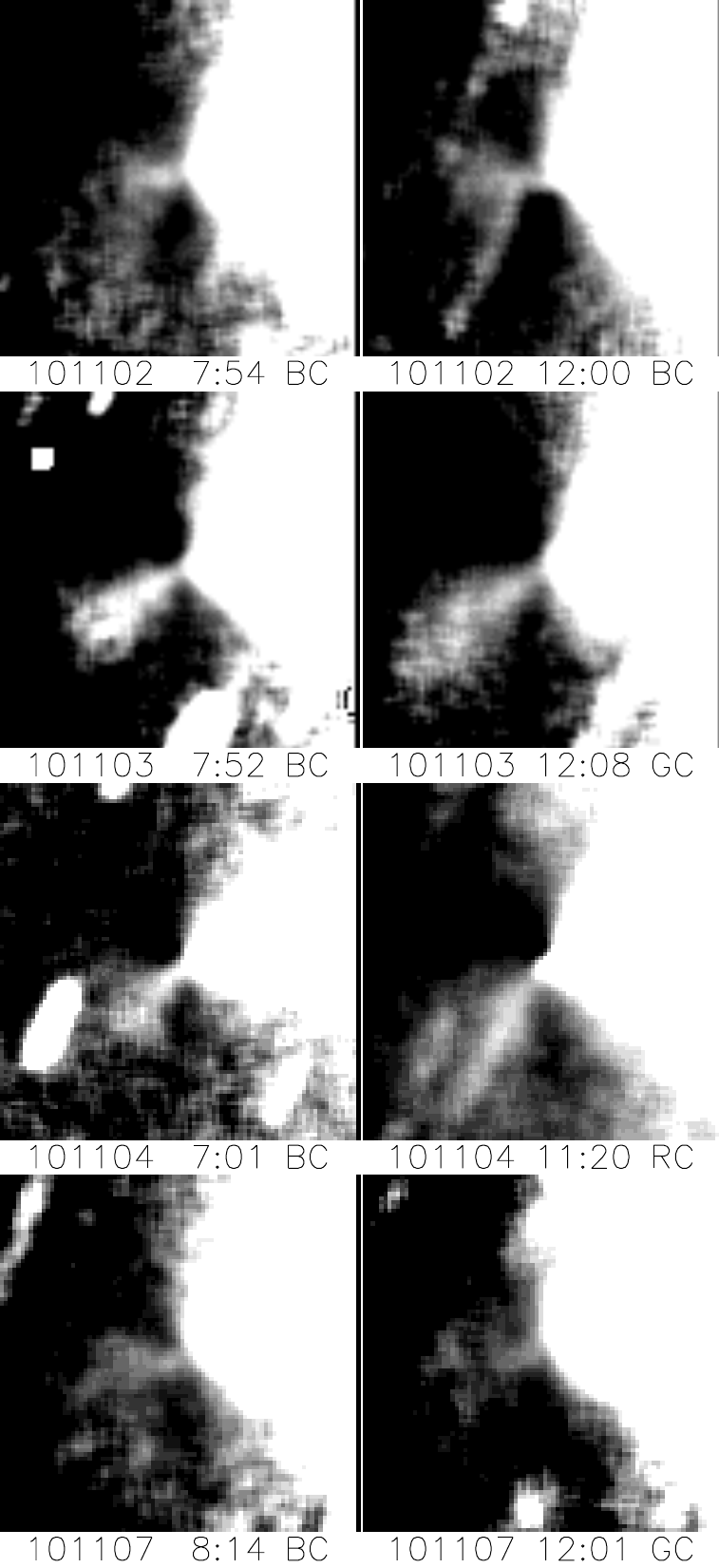}   
  \caption[Dust jet]{Hartley 2 dust jet near the beginning and end of 2010 November 2 (top row), 2010 November 3 (second row), 2010 November 4 (third row), 2010 November 7 (bottom row). Each image is a decontaminated pure dust image except for 2010 November 4 (the night of the EPOXI flyby) when it was not photometric, and the original images are shown after bias removal and flat-fielding. Note that BC images are, by definition, considered to be free of contamination (cf. \citealt{farnham00}). The date (YYMMDD), midpoint UT time (HH:MM), and filter of each image is below the image. Each image is $\sim$8,000 km across, centered on the nucleus, enhanced by division of an azimuthal median profile, and smoothed with a boxcar smooth. North is up, east is left, and the position angle (PA) of the Sun is near 105$^\circ$ (exact PAs for each night are given in Table~\ref{t:imaging_circ}). All images have the same stretch. The dust jet can be seen at PAs of $\sim$95$^\circ$ on November 2 , $\sim$115$^\circ$ on November 3, $\sim$125$^\circ$ on November 4, and $\sim$100$^\circ$ on November 7. Diagonal streaks are trailed stars and the bright white area in the west half of each image is the dust tail.}
  \label{fig:dust_jet}
\end{figure}

We detected a very faint sunward facing dust jet near the nucleus, as also reported by \citet{lara11} and \citet{mueller12}. While difficult to see in Figures~\ref{fig:gas_morph1} and \ref{fig:gas_morph2}, it can be seen in the BC profile in Figure~\ref{fig:profiles}, as the sunward profile is slightly higher than the northern or southern profiles out to $\sim$1600 km. The jet was nearly radial and showed minimal change in shape, position angle, or extent during the course of a night, but varied somewhat from night to night. We show examples of the jet near the start and end of the night in enhanced continuum images on 2010 November 2, 3, 4, and 7 in Figure~\ref{fig:dust_jet}. The jet's appearance was relatively similar on November 2 and 7, pointing nearly due east at PAs of $\sim$95$^\circ$ and $\sim$100$^\circ$, respectively. It had a slightly different appearance on November 3 and 4, when it was to the southeast at PAs of $\sim$115$^\circ$ and $\sim$125$^\circ$, and extended farther than on November 2 or 7. At its maximum extent (on November 3), the dust jet was visible to 20--25 arcsec from the nucleus, although signal-to-noise and our enhancement techniques may have prevented us from seeing it extend farther. Hints of the dust jet were apparent in the October 16 and 17 continuum images, but the extent was much smaller ($<$10 arcsec) than in early November despite better signal-to-noise. The lack of a firm detection in October may have been because the jet was inactive or it may simply have been due to projection effects. The dust jet was not detectable in our data from other months, presumably due to the significantly lower signal-to-noise and larger geocentric distance as compared to 2010 October and November. At no time did we observe a corresponding gas feature.


Surprisingly, the dust jet was easier to detect in our narrowband dust images -- blue continuum (BC), green continuum (GC), and red continuum (RC) -- than in the broadband R images. We applied several additional enhancement techniques to the R-band images in an effort to study the dust jet with more temporal resolution than the continuum images offered. However, we could not consistently detect the dust jet in consecutive R-band images despite the R-band images having a higher signal-to-noise than the continuum images. All of our Hartley 2 observations were focused for the HB narrowband filters, resulting in a typical R-band FWHM of $\sim$6 arcsec. Given the faintness of the dust jet, this R-band defocusing appears to have been sufficient to obscure the dust jet. Since, as was the case for C$_3$, C$_2$, OH, and NH, we only obtained narrowband dust images occasionally, we do not have full rotational coverage. Thus we cannot determine how the dust jet varied with rotation, but our limited data exhibit a slower rate of change than was exhibited by the gas jet(s).

\section{DISCUSSION}
\label{sec:disc}

The coma morphologies varied across the five gas species we observed. At one extreme was CN whose morphology was dominated by an hourglass shaped feature. At the opposite extreme was OH, whose morphology appeared to be an amalgamation of the CN hourglass feature and the tail-dominated dust. In between these extremes were C$_3$ and C$_2$, whose morphologies were generally quite similar to the CN, and the NH, which appeared to be intermediate to the CN and the OH. 

The presence of the hourglass shape exhibited clearly in CN, C$_3$, and C$_2$, and to a lesser extent by NH and OH suggests that all five species originate from the same source region(s) on the nucleus. 
This is supported by the observation that the relative brightness of the northern and southern features in CN, C$_3$, C$_2$, and NH vary in concert, and also that the CN, C$_3$, and C$_2$ morphologies are in agreement from night to night. The relative brightness of the OH hourglass features did not vary as much as the other gas species, and OH also exhibited more near-nucleus coma signal in the tailward direction than the other species. 

The hemispheric asymmetry of OH and NH as well as the relative uniformity of their bulk brightness (after enhancement) suggest that 
a substantial fraction of the OH and NH is derived from small icy grains which survived long enough to be subject to radiation pressure and swept tailward. The infilling of OH and NH throughout the tailward hemisphere is likely due to
the velocities in random directions acquired by the parent molecules as they leave the grains.
\epoxi\ observed numerous individual chunks up to $\sim$20 cm in radius within 30 km of the nucleus \citep{ahearn11a} and \citet{harmon11} reported a significant population of large ($>$cm) dust grains near the nucleus. A'Hearn {\it et al.} suggest that these large grains break up into $\sim$1 $\mu$m solid grains, with the smaller grains providing the surface area necessary to explain Hartley 2's ``hyperactivity.'' 
The bulk brightness enhancement of the OH and NH in the tailward hemisphere supports this model, with the micron-sized grains containing the water and ammonia that eventually produced the OH and NH.

\epoxi\ revealed distinctly different terrain on the nucleus, with a smooth ``waist'' connecting two rougher lobes \citep{ahearn11a}. A'Hearn {\it et al}. determined that different material was coming from these two regions, with H$_2$O vapor coming primarily from the waist and CO$_2$, H$_2$O ice, and organics coming primarily from an end. 
Thus, it is likely that the CN, C$_3$, C$_2$, OH, and NH originate near an end of a lobe, 
presumably from one or more of the many active regions observed by \epoxi. Another very strong constraint is the observation from our photometry that the ratios of the minor species to water are normal; therefore, the vast majority of all of our observed species must be originating at the same time and location.
The differences in the bulk morphology between the NH and OH and the CN, C$_3$, and C$_2$ can be explained if the icy grains containing OH and NH are separated from the parents/grandparents of CN, C$_3$, and C$_2$ (either as smaller grains or simply as vapor) soon after being released from the nucleus. 

Such a scenario suggests that there may have been an intrinsic difference in the protosolar grains out of which Hartley 2 formed, with the parents of OH and NH (primarily water and ammonia, respectively) preferentially being contained in larger grains (and possibly intermixed as ``dirty'' ice) while the parents of CN, C$_2$, and C$_3$ were contained in smaller grains or were deposited primarily on the surface of larger grains. It is somewhat troubling that such a difference in grain compositions has never been detected in other comets (especially  in light of Hartley 2's ``typical'' composition), but given the relative uniqueness of Hartley 2's ``hyperactivity,'' it is not surprising that it may have exhibited other unusual properties; note that no other comet showing hyperactivity has been visited by a spacecraft. The mechanism for lifting large grains, suggested to be CO$_2$ \citep{ahearn11a}, may be vigorous enough to lift off large chunks of the surface {\it in toto}, likely resulting in a considerably different removal of material than the canonical model of cometary activity which envisions gas leaving through pores in the surface and entraining small bits of dust in the process. Conversely, very few comets have been imaged in NH and/or OH owing to their low signal and high atmospheric extinction. Thus, the different coma morphologies of OH and NH compared with the carbon bearing species may not, in fact, be unusual.

As discussed in Section~\ref{sec:dust_morphology}, the dust jet showed little variation during a night but changed somewhat from night to night, always being in the same general direction. These two points are consistent with the apparent non-principal axis rotation of the nucleus (cf. \citealt{ahearn11a,samarasinha11}; Paper 1). The first point suggests that the source of the jet might be located near the comet's total angular momentum vector, the axis around which the nucleus was apparently ``precessing'' with a period near 18.3 hr at the time (cf. \citealt{ahearn11a}), resulting in little change in the morphology over a night. The second point could be caused by Hartley 2's longer ``rotation'' period of $\sim$55 hr at the time of the observations. This would not have much effect on the jet's appearance over the course of one night (only $\sim$10\% of a 55 hr period), but would cause changes from night to night (when the rotational phase had changed by roughly 45\%).

The lack of a gas feature corresponding to the dust jet implies that the dust is lifted from the surface by some other volatile. The most likely candidates are CO and, more likely, CO$_2$, which \citet{ahearn11a} concluded drives activity on Hartley 2. Presumably this source has little to no CN, C$_3$, C$_2$, OH, or NH since no corresponding gas jet was seen; their absence implies that the source has a different composition than the source(s) of the hourglass features and suggests some heterogeneity of the Hartley 2 nucleus. We do not see evidence of radiation pressure affecting the dust jet, which implies we are primarily seeing large dust grains. The velocity dispersion of large grains would be expected to mask much of the rotational signature of the dust jets. Since we do not see any evidence of dust jets near the locations of the hourglass features in the gas, we infer that the gas jet(s) do not have substantial quantities of large grains. Instead, the jet(s) likely lifts small grains which are rapidly pushed tailward; a similar population of small dust grains subject to radiation pressure is likely present in the dust jet and goes similarly undetected.



We next turn to the observation (Section~\ref{sec:composition}) that the production rates have trended downwards steadily from 1991 to 1997/98 to 2010/11
even after accounting for the increase in perihelion distance between apparitions. We speculate that this may be due to the relative youth of the nucleus. In the early twentieth century, Hartley 2 had a perihelion distance ($q$) near 2 AU and an orbital period near 9 yr. Close approaches to Jupiter in 1947 and 1971 caused a drop in the perihelion distance, putting Hartley 2 in an orbit reaching $\sim$1 AU. 
If the primary driver of activity was something more volatile than H$_2$O such as CO$_2$ as suggested by \citet{ahearn11a} (CO is less likely due to the extremely low abundance; \citealt{weaver11}), then the injection into a smaller-$q$ orbit may be rapidly depleting the CO$_2$. Since vigorous activity is required to lift off large chunks of particles necessary to produce the ``hyperactivity'', a rapid decrease in the production of CO$_2$ would cause a correspondingly large drop in the production rate of H$_2$O.

The factor of 1.5--3 drop in the production rates between 1997/98 and 2010/11 spanned two perihelion passages (we are not aware of any published production rate measurements from the 2004 apparition), allowing additional time for production from such areas to decrease. As there were no systematic surveys that would have been sensitive to this intrinsically faint comet on earlier, less favorable, apparitions, we cannot place meaningful constraints on the production rates prior to its discovery in 1986. 
If a similar rate of decrease in production rate occurred between 1971 and 1991, Hartley 2 may have been as much as a factor of 8--10 more active in 1971 (when it was first perturbed into the current orbit) than in 2010/11. 

An alternative, but less likely, explanation for the decreased production rates since 1997/98 may simply be that the illumination of active regions was different between the apparitions.
Hartley 2 exhibited a rapid spin-down coupled with non-principal axis rotation during the 2010/11 apparition (cf. \citealt{ahearn11a,knight11b,samarasinha11}). It is possible that on previous orbits the interplay of the complex rotation and the apparently increasing and decreasing component periods (cf. the Supporting Online Materials from \citealt{ahearn11a}) could result in a different illumination of one or more active regions. This might include the Sun reaching a higher altitude, variations in topography causing less shadowing, the exposure of additional active regions, or the Sun remaining above an active region's local horizon longer. 
However, the nucleus' complex rotation state should minimize seasonal effects; unless the nucleus was very recently (since 1997/98) excited into the present non-principal axis rotation state, large changes in the production rates due entirely to geometry are considered unlikely.


\section{SUMMARY}
\label{sec:summary}
We have presented photometry and imaging of 103P/Hartley 2 obtained at Lowell Observatory (and one night at Perth Observatory) from 1991 through 2011. The photometry includes three apparitions as no data were obtained on the 2004 apparition. We find a secular decrease in brightness from the 1991 apparition to the 1997/98 apparition and then to the 2010/11 apparition and a signficant seasonal effect, with the comet reaching peak brightness $\sim$10 days after perihelion and the rate of brightening steeper than the rate of fading. We find similar results when compiling published water production rates from the literature, consistent with about a 40\% decrease each orbit. Hartley 2's composition is ``typical,'' in agreement with the results of other investigators. We propose two scenarios for the relatively large decrease in production rates: that production rates are dropping rapidly owing to the rapid depletion of CO$_2$ (due to the decrease in perihelion distance in the mid-twentieth century) or, less likely, that the ``complex'' nucleus rotation resulted in progressively less illumination reaching the primary active regions from 1991 to 1997/98 to 2010/11.

Our imaging covered the 2010/11 apparition and focused on the gas coma morphology. We previously reported on our extensive CN data set (Paper 1) and analyze here the morphology exhibited by other gas species (OH, NH, C$_3$, and C$_2$) as well as the dust. We find that C$_3$ and C$_2$ exhibit coma morphology generally similar to what we previously reported for CN, with an hourglass shape in October and November that followed the rotational changes exhibited by CN, and little excess signal in the tailward direction. We conclude that 
differences between the CN, C$_3$, and C$_2$ coma morphologies can be explained by their different lifetimes and parentages. The OH and NH coma morphologies differ from CN, C$_3$, and C$_2$; while OH and NH show evidence of the hourglass shape near the nucleus, they are relatively uniform in brightness (after enhancement) throughout the tailward hemisphere, and the brightness in the tailward hemisphere does not vary appreciably with rotation. We conclude that OH and NH 
are produced from water and ammonia ices that were contained in small grains which shielded the ices long enough to be affected by radiation pressure and driven in the anti-sunward direction. We speculate on possible explanations for why the OH and NH, but not the CN, C$_3$, or C$_2$, behave in this manner, concluding that all five species most likely originate from the same source region(s) near the end of a lobe of the nucleus (where \epoxi\ saw significant active regions) but may have come from aggregates of different sized grains or grains and vapor. Lower velocities and/or a range of velocities associated with a variety of particle sizes would also naturally cause a dilution of the rotational signature of OH and NH as compared to that seen in the carbon bearing species. We detect the faint sunward-facing dust jet reported by \citet{lara11} and \citet{mueller12} in our continuum images. This jet is much smaller than the hourglass shaped gas feature and does not vary appreciably during a night, although it does vary from night to night. No corresponding gas feature was seen at the location of the dust jet, possibly implying that it is driven by CO$_2$. The dust jet may originate from a source near the total angular momentum vector.


The \epoxi\ flyby and supporting observations of Hartley 2 have revealed that it is an unusual comet which is highly active 
relative to its nucleus size, has a population of large grains in the inner coma, and is in non-principal axis rotation with evolving component periods. As one of a small handful of comets to have been visited by a spacecraft, Hartley 2 represents a rare opportunity to link the macroscopic coma morphology and abundances observed remotely with the nucleus shape and active regions observed {\it in situ} via sophisticated modeling. The strongest constraints on such models are likely the direction of the nucleus' long axis as seen by \epoxi\ during the flyby (cf. \citealt{ahearn11a}) and the alignment of the rotational angular momentum at the time of the Arecibo observations \citep{harmon11}. However, any comprehensive jet modeling should also incorporate the constraints 
discussed here and in Paper 1. These include 
the morphological changes during a night (outward motion of the material and the sense of rotation), differences in morphology from night to night (due to the non-principal axis rotation), and evolution of the morphology from month to month (due to the changing viewing geometry).
It is only by satisfying these varied constraints that a model can produce a coherent description of Comet Hartley 2.


\section*{ACKNOWLEDGEMENTS}
\noindent 
We appreciate the work of Anita Cochran and Nalin Samarasinha in carefully refereeing this manuscript and making helpful suggestions to improve the paper. We are indebted to the following people for help in collecting data: Robert Millis (1991); Tony Farnham (1997/1998 Lowell data); Peter Birch (1997 Perth data); Len Bright, Brian Skiff, Larry Wasserman, and Edward Schwieterman (2010/2011 imaging data). We thank Allison Bair for help in producing Table~\ref{t:table5} and for comparisons with preliminary results of the forthcoming comet database paper. Office space was generously provided for MMK by both the University of Maryland Department of Astronomy and Johns Hopkins University Applied Physics Laboratory while he conducted this work. This work has been supported by numerous NASA grants over the years, most recently NNX08AO42G and NNX09AB51G.

\end{singlespace}



\clearpage

\renewcommand{\arraystretch}{0.70}

\begin{deluxetable}{lccrcccccc}  
\tabletypesize{\scriptsize}
\tablecolumns{11}
\tablewidth{0pt} 
\setlength{\tabcolsep}{0.05in}
\tablecaption{Summary of Hartley 2 imaging observations and geometric parameters.\,\tablenotemark{a}}
\tablehead{   
  \colhead{UT}&
  \colhead{UT}&
  \colhead{Telescope}&
  \colhead{$\Delta$T}&
  \colhead{$r_\mathrm{H}$}&
  \colhead{$\Delta$}&
  \colhead{$\theta$\tablenotemark{b}}&
  \colhead{PA Sun\tablenotemark{c}}&
  \colhead{Filters}&
  \colhead{Conditions}\\
  \colhead{Date}&
  \colhead{Range}&
  \colhead{Diam. (m)}&
  \colhead{(days)}&
  \colhead{(AU)}&
  \colhead{(AU)}&
  \colhead{($^\circ$)}&
  \colhead{($^\circ$)}&
  \colhead{}&
  \colhead{}
}
\startdata
2010 Jul 19&10:00--10:22&1.1&$-$100.8&1.68&0.86&29&44&R,CN&Thin cirrus\rule{0cm}{8pt}\\
2010 Aug 13&03:28--11:48&1.1&$-$75.9&1.46&0.57&30&22&R,CN,BC,C3,C2,GC&Photometric\\
2010 Aug 14&03:03--11:48&1.1&$-$74.9&1.45&0.56&30&21&R,CN,BC,C3,C2,GC&Photometric\\
2010 Aug 15&03:15--12:10&1.1&$-$73.9&1.45&0.55&30&20&R,CN,BC&Clouds\\
2010 Aug 16&07:40--11:54&1.1&$-$72.8&1.44&0.54&31&19&R,CN,BC&Clouds\\
2010 Aug 17&07:48--11:01&1.1&$-$71.9&1.43&0.53&31&18&R,CN,BC&Clouds\\
2010 Sep 9&02:53--12:16&1.1&$-$48.9&1.25&0.33&37&357&R,CN,BC,C3,C2,GC,OH&Photometric\\
2010 Sep 10&02:36--12:09&1.1&$-$48.0&1.24&0.32&37&357&R,CN,BC,C3,C2,GC,OH&Photometric\\
2010 Sep 11&02:33--12:08&1.1&$-$47.0&1.24&0.31&37&356&R,CN,BC,C3,C2,GC,OH&Photometric\\
2010 Sep 12&02:33--12:11&1.1&$-$46.0&1.23&0.31&38&356&R,CN,BC&Clouds\\
2010 Sep 13&02:35--12:11&1.1&$-$44.9&1.22&0.30&38&355&R,CN,BC,C3&Clouds\\
2010 Oct 12&03:12--12:34&0.8&$-$15.9&1.08&0.13&49&43&R,CN&Photometric\\
2010 Oct 13&03:13--12:43&0.8&$-$14.9&1.08&0.13&49&48&R,CN&Photometric\\
2010 Oct 14&03:13--12:40&0.8&$-$13.9&1.08&0.13&50&52&R,CN&Photometric\\
2010 Oct 15&03:06--05:33&0.8&$-$13.1&1.07&0.13&50&56&R,CN&Clouds\\
2010 Oct 16&05:01--12:21&1.1&$-$11.9&1.07&0.12&51&61&R,CN,BC,C3,C2,GC,OH,UC,NH&Thin cirrus\\
2010 Oct 17&05:00--12:38&1.1&$-$10.9&1.07&0.12&51&65&R,CN,BC&Clouds\\
2010 Oct 19&10:56--12:24&1.1&$-$8.8&1.07&0.12&52&73&R,CN,BC&Clouds\\
2010 Oct 31&07:10--12:36&0.8&$+$3.2&1.06&0.14&58&99&R,CN&Thin cirrus\\
2010 Nov 1&07:15--12:45&0.8&$+$4.2&1.06&0.14&59&100&R,CN&Thin cirrus\\
2010 Nov 2&06:45--12:54&1.1&$+$5.2&1.06&0.15&59&102&R,CN,BC,C3,C2,GC,OH,UC,NH,RC&Photometric\\
2010 Nov 2&07:27--10:32&0.8&$+$5.1&1.06&0.15&59&102&R,CN&Photometric\\
2010 Nov 3&06:41--13:01&1.1&$+$6.2&1.06&0.15&59&103&R,CN,BC,C3,C2,GC,OH,UC,NH,RC&Photometric\\
2010 Nov 4&06:39--13:07&1.1&$+$7.2&1.06&0.16&59&104&R,CN,BC,C3,C2,GC,OH,UC,H20+,RC&Thin cirrus\\
2010 Nov 5&07:44--10:45&0.8&$+$8.1&1.06&0.16&59&105&R,CN&Photometric\\
2010 Nov 6&07:41--10:58&0.8&$+$9.1&1.07&0.16&59&106&R,CN&Thin cirrus\\
2010 Nov 7&06:48--13:09&1.1&$+$10.2&1.07&0.17&59&107&R,CN,BC,C3,C2,GC,OH,UC,NH,RC&Photometric\\
2010 Nov 10&07:59--13:11&0.8&$+$13.2&1.07&0.18&58&109&R,CN&Clouds\\
2010 Nov 12&07:56--12:30&0.8&$+$15.2&1.08&0.19&58&111&R,CN&Clouds\\
2010 Nov 13&08:09--13:10&0.8&$+$16.2&1.08&0.20&57&112&R,CN&Clouds\\
2010 Nov 16&08:08--13:08&0.8&$+$19.2&1.09&0.21&56&114&R,CN&Photometric\\
2010 Nov 26&07:45--12:43&0.8&$+$29.2&1.13&0.26&51&123&R,CN&Photometric\\
2010 Nov 27&07:48--12:46&0.8&$+$30.2&1.14&0.27&50&124&R,CN&Photometric\\
2010 Dec 9&07:02--13:19&1.1&$+$42.2&1.21&0.33&42&136&R,CN,BC&Thin cirrus\\
2010 Dec 10&06:48--08:50&1.1&$+$43.1&1.21&0.34&41&137&R,CN,BC&Thin cirrus\\
2010 Dec 15&08:57--09:13&1.1&$+$48.1&1.25&0.36&38&143&CN&Clouds\\
2011 Jan 7&04:02--10:51&1.1&$+$71.1&1.42&0.51&25&179&R,CN,BC&Clouds\\
2011 Jan 8&07:12--07:57&1.1&$+$72.1&1.43&0.52&25&181&R,CN&Clouds\\
2011 Jan 9&07:09--08:47&1.1&$+$73.1&1.44&0.53&25&183&R,CN&Clouds\\
2011 Jan 11&04:01--10:50&1.1&$+$75.1&1.45&0.54&24&187&R,CN,BC&Thin cirrus\\

\enddata
\tablenotetext{a} {All parameters are given for the midpoint of each night's observations, and all images were obtained at Lowell Observatory.}
\tablenotetext{b} {Phase angle.}
\tablenotetext{c} {Position angle of the Sun, measured counterclockwise from north through east.}
\label{t:imaging_circ}
\end{deluxetable}

\begin{deluxetable}{lllccccccccc}  
\tabletypesize{\scriptsize}
\tablecolumns{12}
\tablewidth{0pt} 
\setlength{\tabcolsep}{0.03in}
\tablecaption{Photometry observing circumstances and fluorescence efficiencies for Comet 103P/Hartley 2.\,\tablenotemark{a}}
\tablehead{   
  \multicolumn{3}{c}{UT Date}&
  \colhead{$\Delta$T\,\tablenotemark{b}}&
  \colhead{$r_\mathrm{H}$}&
  \colhead{$\Delta$}&
  \colhead{Phase}&
  \colhead{Phase Adj.}&
  \colhead{$\dot{r}_\mathrm{H}$}&
  \multicolumn{3}{c}{log $L/N$ (erg s$^{-1}$ molecule$^{-1}$)}\\
  \cmidrule(){10-12}
  \colhead{}&
  \colhead{}&
  \colhead{}&
  \colhead{(day)}&
  \colhead{(AU)}&
  \colhead{(AU)}&
  \colhead{Angle ($^\circ$)}&
  \colhead{log$A$(0\deg)$f\rho$\tablenotemark{c}\phantom{9}}&
  \colhead{(km s$^{-1}$)}&
  \colhead{OH}&
  \colhead{NH}&
  \colhead{CN}
}
\startdata
1991&Oct&\phantom{0}8.49&\phantom{0}+27.2&1.033&1.046&57.5&+0.47&\phantom{0}+9.7&$-$14.544&$-$13.096&$-$12.351\rule{0cm}{8pt}\\
1991&Oct&11.44&\phantom{0}+30.2&1.050&1.063&56.4&+0.47&+10.5&$-$14.507&$-$13.121&$-$12.370\\[1.0ex]

1997&Nov&\phantom{0}2.13&\phantom{0}$-$49.1&1.241&1.019&50.9&+0.47&$-$12.7&$-$14.735&$-$13.328&$-$12.648\\
1997&Dec&\phantom{0}1.53&\phantom{0}$-$19.7&1.071&0.918&58.8&+0.46&\phantom{0}$-$6.5&$-$14.839&$-$13.177&$-$12.489\\
1997&Dec&\phantom{0}4.16&\phantom{0}$-$17.1&1.062&0.908&59.4&+0.46&\phantom{0}$-$5.8&$-$14.818&$-$13.172&$-$12.495\\
1998&Feb&26.23&\phantom{0}+66.9&1.370&1.091&45.7&+0.46&+14.4&$-$14.504&$-$13.397&$-$12.642\\[1.0ex]

2010&Jul&12.35&$-$107.9&1.742&0.958&29.0&+0.38&$-$15.4&$-$15.215&$-$13.607&$-$12.943\\
2010&Aug&11.30&\phantom{0}$-$78.0&1.480&0.588&29.9&+0.38&$-$14.7&$-$15.011&$-$13.471&$-$12.799\\
2010&Aug&12.24&\phantom{0}$-$77.0&1.472&0.578&30.0&+0.38&$-$14.7&$-$15.006&$-$13.466&$-$12.793\\
2010&Sep&\phantom{0}7.21&\phantom{0}$-$51.1&1.266&0.344&35.8&+0.42&$-$12.4&$-$14.745&$-$13.346&$-$12.670\\
2010&Sep&30.16&\phantom{0}$-$28.1&1.128&0.189&44.4&+0.46&\phantom{0}$-$8.1&$-$14.870&$-$13.220&$-$12.519\\
2010&Oct&\phantom{0}1.17&\phantom{0}$-$27.1&1.123&0.183&44.7&+0.46&\phantom{0}$-$7.9&$-$14.873&$-$13.215&$-$12.516\\
2010&Oct&31.30&\phantom{0}\phantom{0}+3.0&1.060&0.140&58.2&+0.47&\phantom{0}+1.0&$-$14.857&$-$13.223&$-$12.597\\
2010&Nov&16.38&\phantom{0}+19.1&1.091&0.211&56.0&+0.47&\phantom{0}+5.8&$-$14.650&$-$13.135&$-$12.400\\
2010&Dec&13.34&\phantom{0}+46.1&1.232&0.352&39.3&+0.44&+11.6&$-$14.561&$-$13.272&$-$12.527\\
2011&Jan&\phantom{0}5.34&\phantom{0}+69.1&1.405&0.497&26.0&+0.35&+14.2&$-$14.532&$-$13.417&$-$12.668\\
2011&Feb&\phantom{0}1.17&\phantom{0}+95.9&1.635&0.742&21.8&+0.31&+15.3&$-$14.656&$-$13.558&$-$12.793\\
2011&Feb&\phantom{0}2.25&\phantom{0}+97.0&1.645&0.754&21.8&+0.31&+15.3&$-$14.662&$-$13.564&$-$12.799\\
2011&Feb&23.14&+117.9&1.831&1.025&24.2&+0.33&+15.4&$-$14.757&$-$13.658&$-$12.896\\

\enddata
\tablenotetext{a} {All parameters were taken at the midpoint of each night's observations, and all photometry was obtained at Lowell Observatory except for 1997 Dec 1 which was obtained at Perth Observatory.}
\tablenotetext{b} {Time from perihelion.}
\tablenotetext{c} {Adjustment to 0\deg\ phase angle to \afrho\ values based on assumed phase function (see text).}
\label{t:phot_circ}
\end{deluxetable}

\begin{deluxetable}{lllccccccccccccccc}  
\tabletypesize{\scriptsize}
\tablecolumns{18}
\tablewidth{0pt} 
\setlength{\tabcolsep}{0.03in}
\tablecaption{Photometric fluxes and aperture abundances for Comet 103P/Hartley 2}
\tablehead{   
  \colhead{}&
  \colhead{}&
  \colhead{}&
  \multicolumn{2}{c}{Aperture}&
  \multicolumn{5}{c}{log Emission Band Flux}&
  \multicolumn{3}{c}{log Continuum Flux\tablenotemark{a,b}}&
  \multicolumn{5}{c}{log M($\rho$)}\\
  \cmidrule(lr){4-5}
  \colhead{}&
  \colhead{}&
  \colhead{}&
  \colhead{Size}&
  \colhead{log $\rho$}&
  \multicolumn{5}{c}{(erg cm$^{-2}$ s$^{-1}$)}&
  \multicolumn{3}{c}{(erg cm$^{-2}$ s$^{-1}$ \AA$^{-1}$)}&
  \multicolumn{5}{c}{(molecule)}\\
  \cmidrule(lr){6-10}
  \cmidrule(lr){11-13}
  \cmidrule(lr){14-18}
  \multicolumn{3}{c}{UT Date}&
  \colhead{(arcsec)}&
  \colhead{(km)}&
  \colhead{OH}&
  \colhead{NH}&
  \colhead{CN}&
  \colhead{C$_3$}&
  \colhead{C$_2$}&
  \colhead{UV}&
  \colhead{Blue}&
  \colhead{Green}&
  \colhead{OH}&
  \colhead{NH}&
  \colhead{CN}&
  \colhead{C$_3$}&
  \colhead{C$_2$}

}
\startdata
1991&Oct&\phantom{0}8.47&\phantom{0}35.3&4.13&\phantom{0}$-$9.70&$-$10.57&\phantom{0}$-$9.85&$-$10.13&\phantom{0}$-$9.89&\phantom{$<$}$-$13.45&...&$-$13.20&32.33&30.02&29.99&29.38&29.97\rule{0cm}{8pt}\\
1991&Oct&\phantom{0}8.50&\phantom{0}35.3&4.13&\phantom{0}$-$9.70&$-$10.55&\phantom{0}$-$9.85&$-$10.12&\phantom{0}$-$9.88&\phantom{$<$}$-$13.42&...&$-$13.22&32.34&30.03&29.99&29.40&29.98\\
1991&Oct&11.44&\phantom{0}35.3&4.13&\phantom{0}$-$9.78&$-$10.72&\phantom{0}$-$9.99&$-$10.29&$-$10.00&\phantom{$<$}$-$13.45&...&$-$13.27&32.23&29.90&29.88&29.25&29.89\\[1.0ex]
1997&Nov&\phantom{0}2.13&\phantom{0}81.4&4.48&$-$10.37&$-$11.33&$-$10.71&$-$10.81&$-$10.67&\phantom{$<$}$-$14.33&$-$13.87&$-$13.88&31.83&29.47&29.40&28.84&29.33\\
1997&Dec&\phantom{0}1.53&109.9&4.56&...&...&\phantom{0}$-$9.89&$-$10.31&\phantom{0}$-$9.90&...&$-$13.29&...&...&...&29.97&29.12&29.88\\
1997&Dec&\phantom{0}4.16&146.7&4.68&\phantom{0}$-$9.54&$-$10.26&\phantom{0}$-$9.74&$-$10.19&\phantom{0}$-$9.75&\phantom{$<$}$-$13.00&$-$13.06&$-$13.17&32.64&30.28&30.12&29.23&30.01\\
1998&Feb&26.23&114.7&4.66&\phantom{0}$-$9.92&$-$11.24&$-$10.48&$-$10.91&$-$10.51&\phantom{$<$}$-$13.62&$-$13.57&$-$13.81&32.11&29.68&29.69&28.88&29.63\\[1.0ex]
2010&Jul&12.32&\phantom{0}97.2&4.53&$-$11.65&$-$12.55&$-$11.65&$-$11.69&$-$11.79&$<$$-$14.7\phantom{0}&$-$14.39&$-$14.28&30.98&28.47&28.71&28.21&28.45\\
2010&Jul&12.38&\phantom{0}97.2&4.53&$-$11.60&$-$13.74&$-$11.68&$-$11.72&$-$11.76&\phantom{$<$}$-$14.65&$-$14.55&$-$14.44&31.03&27.28&28.68&28.17&28.48\\
2010&Aug&11.30&\phantom{0}97.2&4.32&$-$10.99&$-$11.85&$-$11.15&$-$11.15&$-$11.16&\phantom{$<$}$-$14.36&$-$14.23&$-$14.22&31.01&28.60&28.64&28.18&28.52\\
2010&Aug&11.31&\phantom{0}97.2&4.32&$-$11.01&$-$11.91&$-$11.14&$-$11.15&$-$11.17&\phantom{$<$}$-$14.24&$-$14.09&$-$14.24&30.99&28.55&28.64&28.17&28.51\\
2010&Aug&12.20&155.9&4.51&$-$10.66&$-$11.34&$-$10.86&$-$11.00&$-$10.85&\phantom{$<$}$-$15.13&$-$14.21&$-$15.02&31.32&29.09&28.90&28.31&28.80\\
2010&Aug&12.23&\phantom{0}97.2&4.31&$-$10.98&$-$11.89&$-$11.16&$-$11.20&$-$11.18&{\it und}&$-$14.40&$-$14.28&31.00&28.55&28.61&28.11&28.48\\
2010&Aug&12.26&\phantom{0}97.2&4.31&$-$10.99&$-$11.90&$-$11.20&$-$11.15&$-$11.15&\phantom{$<$}$-$14.43&$-$14.15&$-$14.34&30.99&28.54&28.57&28.16&28.51\\
2010&Aug&12.28&\phantom{0}48.6&4.01&$-$11.49&$-$12.36&$-$11.67&$-$11.59&$-$11.67&\phantom{$<$}$-$14.62&$-$14.58&$-$14.44&30.49&28.08&28.09&27.72&27.98\\
2010&Sep&\phantom{0}7.13&\phantom{0}62.4&3.89&$-$10.60&$-$11.57&$-$10.92&$-$10.76&$-$10.92&\phantom{$<$}$-$14.06&$-$13.71&$-$13.68&30.67&28.29&28.27&27.97&28.15\\
2010&Sep&\phantom{0}7.18&\phantom{0}62.4&3.89&$-$10.60&$-$11.60&$-$10.92&$-$10.72&$-$10.93&\phantom{$<$}$-$14.22&$-$13.69&$-$13.73&30.66&28.27&28.27&28.01&28.14\\
2010&Sep&\phantom{0}7.22&\phantom{0}62.4&3.89&$-$10.61&$-$11.57&$-$10.93&$-$10.74&$-$10.92&\phantom{$<$}$-$14.14&$-$13.71&$-$13.74&30.66&28.30&28.26&27.98&28.15\\
2010&Sep&\phantom{0}7.24&\phantom{0}97.2&4.08&$-$10.29&$-$11.25&$-$10.63&$-$10.50&$-$10.61&\phantom{$<$}$-$14.13&$-$13.53&$-$13.58&30.98&28.62&28.56&28.23&28.46\\
2010&Sep&\phantom{0}7.26&\phantom{0}62.4&3.89&$-$10.62&$-$11.63&$-$10.95&$-$10.78&$-$10.92&\phantom{$<$}$-$14.09&$-$13.71&$-$13.73&30.65&28.24&28.24&27.95&28.15\\
2010&Sep&\phantom{0}7.28&155.9&4.29&\phantom{0}$-$9.96&$-$10.96&$-$10.32&$-$10.45&$-$10.28&{\it und}&$-$13.43&$-$13.82&31.30&28.91&28.88&28.28&28.80\\
2010&Sep&30.11&\phantom{0}62.4&3.63&$-$10.24&$-$11.02&$-$10.46&$-$10.22&$-$10.42&\phantom{$<$}$-$13.54&$-$13.08&$-$13.12&30.63&28.21&28.06&27.89&28.03\\
2010&Sep&30.17&\phantom{0}48.6&3.52&$-$10.43&$-$11.25&$-$10.61&$-$10.37&$-$10.60&\phantom{$<$}$-$13.56&$-$13.17&$-$13.20&30.45&27.97&27.91&27.73&27.85\\
2010&Sep&30.19&155.9&4.03&\phantom{0}$-$9.57&$-$10.36&\phantom{0}$-$9.76&\phantom{0}$-$9.73&\phantom{0}$-$9.76&\phantom{$<$}$-$13.05&$-$12.74&$-$12.81&31.31&28.86&28.76&28.38&28.69\\
2010&Sep&30.20&\phantom{0}97.2&3.82&\phantom{0}$-$9.91&$-$10.71&$-$10.10&\phantom{0}$-$9.97&$-$10.10&\phantom{$<$}$-$13.32&$-$12.94&$-$12.96&30.96&28.52&28.42&28.13&28.35\\
2010&Oct&\phantom{0}1.17&\phantom{0}97.2&3.81&\phantom{0}$-$9.86&$-$10.63&\phantom{0}$-$9.98&\phantom{0}$-$9.84&$-$10.04&\phantom{$<$}$-$13.33&$-$12.88&$-$12.94&30.99&28.56&28.51&28.24&28.38\\
2010&Oct&31.27&\phantom{0}62.4&3.50&\phantom{0}$-$9.90&$-$10.70&$-$10.06&\phantom{0}$-$9.74&\phantom{0}$-$9.94&\phantom{$<$}$-$13.14&$-$12.66&$-$12.72&30.70&28.26&28.28&28.05&28.20\\
2010&Oct&31.28&126.7&3.81&\phantom{0}$-$9.30&$-$10.12&\phantom{0}$-$9.53&\phantom{0}$-$9.32&\phantom{0}$-$9.41&\phantom{$<$}$-$12.80&$-$12.40&$-$12.46&31.29&28.84&28.80&28.47&28.73\\
2010&Oct&31.30&\phantom{0}48.6&3.39&$-$10.04&$-$10.87&$-$10.25&\phantom{0}$-$9.90&$-$10.13&\phantom{$<$}$-$13.23&$-$12.80&$-$12.83&30.56&28.10&28.09&27.89&28.01\\
2010&Oct&31.30&\phantom{0}38.5&3.29&$-$10.20&$-$11.06&$-$10.42&$-$10.06&$-$10.30&\phantom{$<$}$-$13.35&$-$12.89&$-$12.91&30.40&27.91&27.92&27.74&27.84\\
2010&Oct&31.32&\phantom{0}97.2&3.69&\phantom{0}$-$9.54&$-$10.34&\phantom{0}$-$9.75&\phantom{0}$-$9.49&\phantom{0}$-$9.62&\phantom{$<$}$-$12.95&$-$12.52&$-$12.55&31.06&28.62&28.59&28.30&28.52\\
2010&Nov&16.31&\phantom{0}97.2&3.87&\phantom{0}$-$9.47&$-$10.36&\phantom{0}$-$9.63&\phantom{0}$-$9.69&\phantom{0}$-$9.74&\phantom{$<$}$-$12.94&$-$12.66&$-$12.71&31.27&28.87&28.87&28.48&28.78\\
2010&Nov&16.32&\phantom{0}62.4&3.68&\phantom{0}$-$9.80&$-$10.72&\phantom{0}$-$9.94&\phantom{0}$-$9.91&$-$10.06&\phantom{$<$}$-$13.17&$-$12.87&$-$12.91&30.94&28.51&28.55&28.26&28.46\\
2010&Nov&16.35&\phantom{0}38.5&3.47&$-$10.15&$-$11.06&$-$10.28&$-$10.22&$-$10.42&\phantom{$<$}$-$13.41&$-$13.08&$-$13.12&30.60&28.17&28.22&27.95&28.10\\
2010&Nov&16.39&\phantom{0}77.8&3.77&\phantom{0}$-$9.63&$-$10.49&\phantom{0}$-$9.77&\phantom{0}$-$9.79&\phantom{0}$-$9.88&\phantom{$<$}$-$13.01&$-$12.77&$-$12.82&31.12&28.74&28.73&28.39&28.64\\
2010&Nov&16.40&\phantom{0}97.2&3.87&\phantom{0}$-$9.48&$-$10.34&\phantom{0}$-$9.63&\phantom{0}$-$9.65&\phantom{0}$-$9.74&\phantom{$<$}$-$13.08&$-$12.69&$-$12.75&31.27&28.89&28.86&28.52&28.78\\
2010&Nov&16.41&\phantom{0}48.6&3.57&\phantom{0}$-$9.98&$-$10.85&$-$10.11&$-$10.03&$-$10.22&\phantom{$<$}$-$13.37&$-$12.96&$-$13.01&30.76&28.38&28.39&28.15&28.30\\
2010&Nov&16.42&\phantom{0}62.4&3.68&\phantom{0}$-$9.80&$-$10.68&\phantom{0}$-$9.94&\phantom{0}$-$9.90&$-$10.06&\phantom{$<$}$-$13.31&$-$12.87&$-$12.91&30.95&28.55&28.56&28.27&28.46\\
2010&Nov&16.43&126.7&3.99&\phantom{0}$-$9.29&$-$10.15&\phantom{0}$-$9.45&\phantom{0}$-$9.53&\phantom{0}$-$9.55&\phantom{$<$}$-$12.96&$-$12.59&$-$12.66&31.45&29.08&29.05&28.64&28.97\\
2010&Nov&16.45&\phantom{0}77.8&3.77&\phantom{0}$-$9.63&$-$10.50&\phantom{0}$-$9.76&\phantom{0}$-$9.77&\phantom{0}$-$9.87&\phantom{$<$}$-$13.19&$-$12.77&$-$12.84&31.12&28.74&28.73&28.40&28.65\\
2010&Dec&13.27&\phantom{0}38.5&3.69&$-$10.38&$-$11.42&$-$10.65&$-$10.61&$-$10.81&\phantom{$<$}$-$13.87&$-$13.41&$-$13.42&30.72&28.39&28.42&28.12&28.26\\
2010&Dec&13.28&\phantom{0}48.6&3.79&$-$10.19&$-$11.28&$-$10.47&$-$10.48&$-$10.63&\phantom{$<$}$-$13.66&$-$13.32&$-$13.29&30.91&28.53&28.60&28.25&28.44\\
2010&Dec&13.34&\phantom{0}77.8&4.00&\phantom{0}$-$9.86&$-$10.91&$-$10.16&$-$10.21&$-$10.26&\phantom{$<$}$-$13.48&$-$13.09&$-$13.11&31.24&28.91&28.91&28.51&28.81\\
2010&Dec&13.35&\phantom{0}24.5&3.50&$-$10.70&$-$11.77&$-$10.96&$-$10.88&$-$11.14&\phantom{$<$}$-$14.12&$-$13.59&$-$13.61&30.40&28.04&28.11&27.84&27.93\\
2010&Dec&13.36&\phantom{0}48.6&3.79&$-$10.19&$-$11.28&$-$10.47&$-$10.43&$-$10.61&\phantom{$<$}$-$13.74&$-$13.29&$-$13.34&30.92&28.53&28.60&28.29&28.46\\
2010&Dec&13.37&\phantom{0}97.2&4.09&\phantom{0}$-$9.71&$-$10.81&$-$10.02&$-$10.09&$-$10.12&\phantom{$<$}$-$13.34&$-$13.03&$-$13.10&31.40&29.01&29.05&28.64&28.95\\
2010&Dec&13.38&126.7&4.21&\phantom{0}$-$9.54&$-$10.60&\phantom{0}$-$9.86&\phantom{0}$-$9.98&\phantom{0}$-$9.94&\phantom{$<$}$-$13.20&$-$12.92&$-$13.00&31.57&29.22&29.21&28.74&29.13\\
2010&Dec&13.40&\phantom{0}62.4&3.90&$-$10.02&$-$11.10&$-$10.31&$-$10.32&$-$10.43&\phantom{$<$}$-$13.56&$-$13.24&$-$13.28&31.08&28.71&28.76&28.40&28.64\\
2010&Dec&13.41&\phantom{0}77.8&4.00&\phantom{0}$-$9.86&$-$10.90&$-$10.15&$-$10.18&$-$10.25&\phantom{$<$}$-$13.52&$-$13.14&$-$13.20&31.24&28.91&28.92&28.54&28.82\\
2011&Jan&\phantom{0}5.30&\phantom{0}62.4&4.05&$-$10.39&$-$11.69&$-$10.88&$-$10.93&$-$10.94&\phantom{$<$}$-$14.12&$-$13.64&$-$13.65&30.99&28.57&28.63&28.21&28.54\\
2011&Jan&\phantom{0}5.31&\phantom{0}38.5&3.84&$-$10.71&$-$12.08&$-$11.24&$-$11.19&$-$11.33&\phantom{$<$}$-$14.14&$-$13.84&$-$13.73&30.66&28.17&28.26&27.95&28.15\\
2011&Jan&\phantom{0}5.36&\phantom{0}77.8&4.15&$-$10.20&$-$11.55&$-$10.74&$-$10.84&$-$10.78&\phantom{$<$}$-$13.78&$-$13.47&$-$13.49&31.17&28.71&28.77&28.30&28.71\\
2011&Jan&\phantom{0}5.37&\phantom{0}97.2&4.24&$-$10.06&$-$11.39&$-$10.59&$-$10.69&$-$10.64&\phantom{$<$}$-$13.81&$-$13.43&$-$13.45&31.31&28.87&28.92&28.44&28.84\\
2011&Feb&\phantom{0}1.13&\phantom{0}62.4&4.23&$-$10.91&$-$12.46&$-$11.29&$-$11.43&$-$11.40&\phantom{$<$}$-$14.35&$-$14.02&$-$14.15&30.94&28.29&28.69&28.19&28.56\\
2011&Feb&\phantom{0}1.14&\phantom{0}97.2&4.42&$-$10.57&$-$11.82&$-$10.99&$-$11.09&$-$11.13&\phantom{$<$}$-$14.08&$-$13.68&$-$13.62&31.28&28.92&29.00&28.53&28.83\\
2011&Feb&\phantom{0}1.21&\phantom{0}97.2&4.42&$-$10.59&$-$11.94&$-$11.01&$-$11.11&$-$11.07&\phantom{$<$}$-$14.13&$-$13.83&$-$13.96&31.25&28.81&28.97&28.51&28.90\\
2011&Feb&\phantom{0}2.23&126.7&4.54&$-$10.42&$-$11.95&$-$10.88&$-$11.01&$-$10.96&\phantom{$<$}$-$14.30&$-$13.79&$-$13.81&31.45&28.81&29.13&28.63&29.02\\
2011&Feb&\phantom{0}2.26&\phantom{0}48.6&4.12&$-$11.11&$-$12.96&$-$11.58&$-$11.64&$-$11.71&\phantom{$<$}$-$14.53&$-$14.04&$-$14.16&30.75&27.81&28.42&28.00&28.27\\
2011&Feb&23.11&\phantom{0}62.4&4.37&$-$11.34&$-$12.58&$-$11.74&$-$11.63&$-$11.83&\phantom{$<$}$-$14.72&$-$14.22&$-$14.35&30.89&28.55&28.63&28.37&28.51\\
2011&Feb&23.13&\phantom{0}77.8&4.46&$-$11.19&$-$13.05&$-$11.55&$-$11.58&$-$11.63&\phantom{$<$}$-$14.42&$-$14.22&$-$14.26&31.03&28.08&28.82&28.42&28.71\\
2011&Feb&23.15&\phantom{0}97.2&4.56&$-$11.00&$-$12.34&$-$11.39&$-$11.48&$-$11.51&\phantom{$<$}$-$14.43&$-$13.82&$-$13.95&31.23&28.79&28.98&28.52&28.84\\
2011&Feb&23.18&\phantom{0}62.4&4.37&$-$11.32&$-$12.78&$-$11.67&$-$11.66&$-$11.89&\phantom{$<$}$-$14.75&$-$14.21&$-$14.06&30.91&28.34&28.70&28.33&28.45\\

\enddata
\tablenotetext{a} {Continuum filter wavelengths: UV (1991) = 3650 \AA; UV (1997/98 \& 2010/11) = 3448 \AA; blue = 44450 \AA; green (1991) = 4845 \AA; green (1997/98 \& 2010/11) = 5260 \AA.}
\tablenotetext{b} {``und'' stands for ``undefined'' and means the continuum flux was measured but was less than 0.}
\label{t:phot_flux}
\end{deluxetable}

\begin{deluxetable}{lllcccccccccccc}  
\tabletypesize{\scriptsize}
\tablecolumns{15}
\tablewidth{0pt} 
\setlength{\tabcolsep}{0.03in}
\tablecaption{Photometric production rates for Comet 103P/Hartley 2.}
\tablehead{   
  \colhead{}&
  \colhead{}&
  \colhead{}&
  \colhead{$\Delta$T}&
  \colhead{log $r_\mathrm{H}$}&
  \colhead{log $\rho$}&
  \multicolumn{5}{c}{log $Q$\tablenotemark{a}\phantom{00}(molecules s$^{-1}$)}&
  \multicolumn{3}{c}{log $A$($\theta$)$f\rho$\tablenotemark{a,b,c}\phantom{0,00}(cm)}&
  \colhead{log $Q$}\\
  \cmidrule(lr){7-11}
  \cmidrule(lr){12-14}
  \cmidrule(){15-15}
  \multicolumn{3}{c}{UT Date}&
  \colhead{(day)}&
  \colhead{(AU)}&
  \colhead{(km)}&
  \colhead{OH}&
  \colhead{NH}&
  \colhead{CN}&
  \colhead{C$_3$}&
  \colhead{C$_2$}&
  \colhead{UV}&
  \colhead{Blue}&
  \colhead{Green}&
  \colhead{H$_2$O}

}
\startdata
1991&Oct&\phantom{0}8.47&\phantom{0}+27.2&0.014&4.13&28.50{\tiny\phantom{.}.01}&26.40{\tiny\phantom{.}.01}&25.99{\tiny\phantom{.}.00}&25.31{\tiny\phantom{.}.00}&26.17{\tiny\phantom{.}.00}&2.40{\tiny\phantom{.}.02}&...&2.41{\tiny\phantom{.}.01}&28.62\rule{0cm}{8pt}\\
1991&Oct&\phantom{0}8.50&\phantom{0}+27.2&0.014&4.13&28.50{\tiny\phantom{.}.01}&26.42{\tiny\phantom{.}.01}&25.99{\tiny\phantom{.}.00}&25.33{\tiny\phantom{.}.00}&26.18{\tiny\phantom{.}.00}&2.43{\tiny\phantom{.}.01}&...&2.39{\tiny\phantom{.}.01}&28.63\\
1991&Oct&11.44&\phantom{0}+30.2&0.021&4.13&28.39{\tiny\phantom{.}.01}&26.28{\tiny\phantom{.}.01}&25.88{\tiny\phantom{.}.00}&25.17{\tiny\phantom{.}.01}&26.08{\tiny\phantom{.}.00}&2.42{\tiny\phantom{.}.02}&...&2.36{\tiny\phantom{.}.01}&28.51\\[1.0ex]
1997&Nov&\phantom{0}2.13&\phantom{0}$-$49.1&0.094&4.48&27.55{\tiny\phantom{.}.01}&25.39{\tiny\phantom{.}.02}&24.98{\tiny\phantom{.}.01}&24.46{\tiny\phantom{.}.02}&25.10{\tiny\phantom{.}.01}&1.39{\tiny\phantom{.}.13}&1.52{\tiny\phantom{.}.04}&1.53{\tiny\phantom{.}.03}&27.64\\
1997&Dec&\phantom{0}1.53&\phantom{0}$-$19.7&0.030&4.56&...&...&25.42{\tiny\phantom{.}.00}&24.75{\tiny\phantom{.}.01}&25.52{\tiny\phantom{.}.00}&...&1.80{\tiny\phantom{.}.02}&...&...\\
1997&Dec&\phantom{0}4.16&\phantom{0}$-$17.1&0.026&4.68&28.06{\tiny\phantom{.}.01}&25.85{\tiny\phantom{.}.00}&25.44{\tiny\phantom{.}.00}&24.82{\tiny\phantom{.}.01}&25.52{\tiny\phantom{.}.00}&2.28{\tiny\phantom{.}.02}&1.89{\tiny\phantom{.}.01}&1.80{\tiny\phantom{.}.01}&28.18\\
1998&Feb&26.23&\phantom{0}+66.9&0.137&4.66&27.62{\tiny\phantom{.}.02}&25.38{\tiny\phantom{.}.04}&25.06{\tiny\phantom{.}.01}&24.36{\tiny\phantom{.}.04}&25.20{\tiny\phantom{.}.01}&2.07{\tiny\phantom{.}.09}&1.79{\tiny\phantom{.}.05}&1.57{\tiny\phantom{.}.05}&27.69\\[1.0ex]
2010&Jul&12.32&$-$107.9&0.241&4.53&26.77{\tiny\phantom{.}.09}&24.48{\tiny\phantom{.}.21}&24.33{\tiny\phantom{.}.04}&23.73{\tiny\phantom{.}.12}&24.27{\tiny\phantom{.}.06}&$<$1.2&1.19{\tiny\phantom{.}.14}&1.32{\tiny\phantom{.}.11}&26.78\\
2010&Jul&12.38&$-$107.9&0.241&4.53&26.82{\tiny\phantom{.}.07}&23.30{\tiny\phantom{.}.89}&24.30{\tiny\phantom{.}.03}&23.69{\tiny\phantom{.}.11}&24.30{\tiny\phantom{.}.06}&1.26{\tiny\phantom{.}.28}&1.04{\tiny\phantom{.}.18}&1.16{\tiny\phantom{.}.14}&26.83\\
2010&Aug&11.30&\phantom{0}$-$78.0&0.170&4.32&27.05{\tiny\phantom{.}.02}&24.87{\tiny\phantom{.}.05}&24.50{\tiny\phantom{.}.01}&23.90{\tiny\phantom{.}.04}&24.58{\tiny\phantom{.}.02}&1.19{\tiny\phantom{.}.16}&1.00{\tiny\phantom{.}.10}&1.03{\tiny\phantom{.}.09}&27.10\\
2010&Aug&11.31&\phantom{0}$-$78.0&0.170&4.32&27.03{\tiny\phantom{.}.02}&24.82{\tiny\phantom{.}.05}&24.51{\tiny\phantom{.}.01}&23.90{\tiny\phantom{.}.04}&24.57{\tiny\phantom{.}.02}&1.32{\tiny\phantom{.}.12}&1.13{\tiny\phantom{.}.07}&1.01{\tiny\phantom{.}.09}&27.08\\
2010&Aug&12.20&\phantom{0}$-$77.1&0.168&4.51&27.06{\tiny\phantom{.}.02}&25.04{\tiny\phantom{.}.03}&24.49{\tiny\phantom{.}.08}&23.86{\tiny\phantom{.}.08}&24.58{\tiny\phantom{.}.01}&0.20{\tiny\phantom{.}.79}&0.80{\tiny\phantom{.}.15}&0.01{\tiny\phantom{.}.44}&27.11\\
2010&Aug&12.23&\phantom{0}$-$77.0&0.168&4.31&27.05{\tiny\phantom{.}.02}&24.83{\tiny\phantom{.}.06}&24.48{\tiny\phantom{.}.02}&23.84{\tiny\phantom{.}.05}&24.55{\tiny\phantom{.}.02}&{\it und}&0.81{\tiny\phantom{.}.15}&0.96{\tiny\phantom{.}.11}&27.10\\
2010&Aug&12.26&\phantom{0}$-$77.0&0.168&4.31&27.04{\tiny\phantom{.}.02}&24.82{\tiny\phantom{.}.05}&24.45{\tiny\phantom{.}.02}&23.89{\tiny\phantom{.}.04}&24.58{\tiny\phantom{.}.02}&1.11{\tiny\phantom{.}.17}&1.06{\tiny\phantom{.}.08}&0.90{\tiny\phantom{.}.11}&27.09\\
2010&Aug&12.28&\phantom{0}$-$77.0&0.168&4.01&27.03{\tiny\phantom{.}.03}&24.86{\tiny\phantom{.}.07}&24.44{\tiny\phantom{.}.02}&23.79{\tiny\phantom{.}.05}&24.53{\tiny\phantom{.}.03}&1.23{\tiny\phantom{.}.14}&0.94{\tiny\phantom{.}.11}&1.10{\tiny\phantom{.}.08}&27.08\\
2010&Sep&\phantom{0}7.13&\phantom{0}$-$51.1&0.102&3.89&27.31{\tiny\phantom{.}.01}&25.19{\tiny\phantom{.}.03}&24.72{\tiny\phantom{.}.01}&24.14{\tiny\phantom{.}.02}&24.80{\tiny\phantom{.}.01}&1.32{\tiny\phantom{.}.09}&1.34{\tiny\phantom{.}.03}&1.39{\tiny\phantom{.}.02}&27.40\\
2010&Sep&\phantom{0}7.18&\phantom{0}$-$51.1&0.102&3.89&27.31{\tiny\phantom{.}.01}&25.16{\tiny\phantom{.}.03}&24.72{\tiny\phantom{.}.01}&24.19{\tiny\phantom{.}.01}&24.80{\tiny\phantom{.}.01}&1.16{\tiny\phantom{.}.10}&1.36{\tiny\phantom{.}.02}&1.34{\tiny\phantom{.}.03}&27.39\\
2010&Sep&\phantom{0}7.22&\phantom{0}$-$51.0&0.102&3.89&27.30{\tiny\phantom{.}.01}&25.19{\tiny\phantom{.}.02}&24.71{\tiny\phantom{.}.01}&24.16{\tiny\phantom{.}.01}&24.81{\tiny\phantom{.}.01}&1.24{\tiny\phantom{.}.09}&1.35{\tiny\phantom{.}.03}&1.34{\tiny\phantom{.}.03}&27.39\\
2010&Sep&\phantom{0}7.24&\phantom{0}$-$51.0&0.102&4.08&27.31{\tiny\phantom{.}.01}&25.18{\tiny\phantom{.}.02}&24.71{\tiny\phantom{.}.01}&24.19{\tiny\phantom{.}.01}&24.81{\tiny\phantom{.}.01}&1.06{\tiny\phantom{.}.12}&1.33{\tiny\phantom{.}.02}&1.30{\tiny\phantom{.}.03}&27.39\\
2010&Sep&\phantom{0}7.26&\phantom{0}$-$51.0&0.102&3.89&27.29{\tiny\phantom{.}.01}&25.13{\tiny\phantom{.}.03}&24.69{\tiny\phantom{.}.01}&24.13{\tiny\phantom{.}.01}&24.80{\tiny\phantom{.}.01}&1.29{\tiny\phantom{.}.07}&1.34{\tiny\phantom{.}.02}&1.34{\tiny\phantom{.}.02}&27.37\\
2010&Sep&\phantom{0}7.28&\phantom{0}$-$51.0&0.102&4.29&27.31{\tiny\phantom{.}.01}&25.14{\tiny\phantom{.}.04}&24.72{\tiny\phantom{.}.02}&24.04{\tiny\phantom{.}.04}&24.83{\tiny\phantom{.}.00}&{\it und}&1.22{\tiny\phantom{.}.03}&0.85{\tiny\phantom{.}.07}&27.39\\
2010&Sep&30.11&\phantom{0}$-$28.2&0.052&3.63&27.64{\tiny\phantom{.}.01}&25.47{\tiny\phantom{.}.01}&24.87{\tiny\phantom{.}.01}&24.37{\tiny\phantom{.}.01}&25.05{\tiny\phantom{.}.00}&1.48{\tiny\phantom{.}.03}&1.61{\tiny\phantom{.}.01}&1.59{\tiny\phantom{.}.01}&27.75\\
2010&Sep&30.17&\phantom{0}$-$28.1&0.052&3.52&27.64{\tiny\phantom{.}.01}&25.43{\tiny\phantom{.}.01}&24.90{\tiny\phantom{.}.00}&24.37{\tiny\phantom{.}.01}&25.05{\tiny\phantom{.}.01}&1.57{\tiny\phantom{.}.03}&1.63{\tiny\phantom{.}.01}&1.62{\tiny\phantom{.}.01}&27.75\\
2010&Sep&30.19&\phantom{0}$-$28.1&0.052&4.03&27.66{\tiny\phantom{.}.00}&25.45{\tiny\phantom{.}.01}&24.94{\tiny\phantom{.}.00}&24.39{\tiny\phantom{.}.00}&25.07{\tiny\phantom{.}.00}&1.57{\tiny\phantom{.}.02}&1.55{\tiny\phantom{.}.01}&1.50{\tiny\phantom{.}.01}&27.77\\
2010&Sep&30.20&\phantom{0}$-$28.1&0.052&3.82&27.65{\tiny\phantom{.}.00}&25.45{\tiny\phantom{.}.01}&24.92{\tiny\phantom{.}.00}&24.37{\tiny\phantom{.}.01}&25.05{\tiny\phantom{.}.00}&1.51{\tiny\phantom{.}.03}&1.56{\tiny\phantom{.}.01}&1.56{\tiny\phantom{.}.01}&27.76\\
2010&Oct&\phantom{0}1.17&\phantom{0}$-$27.1&0.050&3.81&27.70{\tiny\phantom{.}.00}&25.52{\tiny\phantom{.}.01}&25.03{\tiny\phantom{.}.00}&24.50{\tiny\phantom{.}.00}&25.10{\tiny\phantom{.}.00}&1.48{\tiny\phantom{.}.03}&1.60{\tiny\phantom{.}.01}&1.57{\tiny\phantom{.}.01}&27.81\\
2010&Oct&31.27&\phantom{0}\phantom{0}+3.0&0.025&3.50&27.89{\tiny\phantom{.}.02}&25.71{\tiny\phantom{.}.01}&25.27{\tiny\phantom{.}.00}&24.69{\tiny\phantom{.}.00}&25.40{\tiny\phantom{.}.00}&1.70{\tiny\phantom{.}.03}&1.85{\tiny\phantom{.}.01}&1.81{\tiny\phantom{.}.01}&28.02\\
2010&Oct&31.28&\phantom{0}\phantom{0}+3.0&0.025&3.81&27.97{\tiny\phantom{.}.01}&25.76{\tiny\phantom{.}.01}&25.30{\tiny\phantom{.}.00}&24.72{\tiny\phantom{.}.00}&25.42{\tiny\phantom{.}.00}&1.73{\tiny\phantom{.}.03}&1.80{\tiny\phantom{.}.01}&1.76{\tiny\phantom{.}.01}&28.10\\
2010&Oct&31.30&\phantom{0}\phantom{0}+3.0&0.025&3.39&27.94{\tiny\phantom{.}.01}&25.74{\tiny\phantom{.}.01}&25.27{\tiny\phantom{.}.00}&24.69{\tiny\phantom{.}.01}&25.39{\tiny\phantom{.}.00}&1.71{\tiny\phantom{.}.03}&1.82{\tiny\phantom{.}.01}&1.81{\tiny\phantom{.}.01}&28.07\\
2010&Oct&31.30&\phantom{0}\phantom{0}+3.0&0.025&3.29&27.95{\tiny\phantom{.}.01}&25.73{\tiny\phantom{.}.01}&25.27{\tiny\phantom{.}.01}&24.68{\tiny\phantom{.}.01}&25.39{\tiny\phantom{.}.00}&1.70{\tiny\phantom{.}.03}&1.83{\tiny\phantom{.}.01}&1.83{\tiny\phantom{.}.01}&28.07\\
2010&Oct&31.32&\phantom{0}\phantom{0}+3.1&0.025&3.69&27.93{\tiny\phantom{.}.00}&25.74{\tiny\phantom{.}.01}&25.26{\tiny\phantom{.}.00}&24.69{\tiny\phantom{.}.00}&25.40{\tiny\phantom{.}.00}&1.69{\tiny\phantom{.}.02}&1.80{\tiny\phantom{.}.01}&1.78{\tiny\phantom{.}.01}&28.05\\
2010&Nov&16.31&\phantom{0}+19.0&0.038&3.87&27.87{\tiny\phantom{.}.01}&25.70{\tiny\phantom{.}.01}&25.27{\tiny\phantom{.}.00}&24.66{\tiny\phantom{.}.01}&25.38{\tiny\phantom{.}.00}&1.90{\tiny\phantom{.}.06}&1.85{\tiny\phantom{.}.02}&1.83{\tiny\phantom{.}.02}&27.98\\
2010&Nov&16.32&\phantom{0}+19.1&0.038&3.68&27.85{\tiny\phantom{.}.01}&25.67{\tiny\phantom{.}.02}&25.27{\tiny\phantom{.}.00}&24.67{\tiny\phantom{.}.01}&25.38{\tiny\phantom{.}.00}&1.87{\tiny\phantom{.}.06}&1.84{\tiny\phantom{.}.02}&1.83{\tiny\phantom{.}.01}&27.97\\
2010&Nov&16.35&\phantom{0}+19.1&0.038&3.47&27.87{\tiny\phantom{.}.01}&25.70{\tiny\phantom{.}.01}&25.28{\tiny\phantom{.}.00}&24.65{\tiny\phantom{.}.01}&25.37{\tiny\phantom{.}.01}&1.84{\tiny\phantom{.}.04}&1.85{\tiny\phantom{.}.01}&1.82{\tiny\phantom{.}.01}&27.98\\
2010&Nov&16.39&\phantom{0}+19.1&0.038&3.77&27.87{\tiny\phantom{.}.00}&25.74{\tiny\phantom{.}.00}&25.29{\tiny\phantom{.}.00}&24.68{\tiny\phantom{.}.00}&25.40{\tiny\phantom{.}.00}&1.93{\tiny\phantom{.}.01}&1.85{\tiny\phantom{.}.01}&1.82{\tiny\phantom{.}.01}&27.99\\
2010&Nov&16.40&\phantom{0}+19.1&0.038&3.87&27.86{\tiny\phantom{.}.00}&25.73{\tiny\phantom{.}.00}&25.27{\tiny\phantom{.}.00}&24.70{\tiny\phantom{.}.00}&25.39{\tiny\phantom{.}.00}&1.76{\tiny\phantom{.}.02}&1.83{\tiny\phantom{.}.01}&1.79{\tiny\phantom{.}.01}&27.98\\
2010&Nov&16.41&\phantom{0}+19.1&0.038&3.57&27.86{\tiny\phantom{.}.00}&25.73{\tiny\phantom{.}.01}&25.28{\tiny\phantom{.}.00}&24.70{\tiny\phantom{.}.00}&25.40{\tiny\phantom{.}.00}&1.78{\tiny\phantom{.}.02}&1.86{\tiny\phantom{.}.01}&1.83{\tiny\phantom{.}.01}&27.97\\
2010&Nov&16.42&\phantom{0}+19.2&0.038&3.68&27.86{\tiny\phantom{.}.00}&25.71{\tiny\phantom{.}.01}&25.28{\tiny\phantom{.}.00}&24.68{\tiny\phantom{.}.00}&25.38{\tiny\phantom{.}.00}&1.73{\tiny\phantom{.}.02}&1.84{\tiny\phantom{.}.01}&1.83{\tiny\phantom{.}.01}&27.97\\
2010&Nov&16.43&\phantom{0}+19.2&0.038&3.99&27.86{\tiny\phantom{.}.00}&25.72{\tiny\phantom{.}.00}&25.28{\tiny\phantom{.}.00}&24.70{\tiny\phantom{.}.00}&25.40{\tiny\phantom{.}.00}&1.77{\tiny\phantom{.}.02}&1.82{\tiny\phantom{.}.01}&1.77{\tiny\phantom{.}.01}&27.98\\
2010&Nov&16.45&\phantom{0}+19.2&0.038&3.77&27.87{\tiny\phantom{.}.00}&25.73{\tiny\phantom{.}.00}&25.29{\tiny\phantom{.}.00}&24.70{\tiny\phantom{.}.00}&25.41{\tiny\phantom{.}.00}&1.75{\tiny\phantom{.}.02}&1.84{\tiny\phantom{.}.01}&1.80{\tiny\phantom{.}.01}&27.99\\
2010&Dec&13.27&\phantom{0}+46.0&0.091&3.69&27.68{\tiny\phantom{.}.03}&25.61{\tiny\phantom{.}.04}&25.18{\tiny\phantom{.}.01}&24.55{\tiny\phantom{.}.02}&25.22{\tiny\phantom{.}.01}&1.71{\tiny\phantom{.}.11}&1.84{\tiny\phantom{.}.03}&1.84{\tiny\phantom{.}.02}&27.77\\
2010&Dec&13.28&\phantom{0}+46.0&0.091&3.79&27.71{\tiny\phantom{.}.02}&25.58{\tiny\phantom{.}.02}&25.19{\tiny\phantom{.}.01}&24.54{\tiny\phantom{.}.01}&25.24{\tiny\phantom{.}.01}&1.82{\tiny\phantom{.}.06}&1.83{\tiny\phantom{.}.02}&1.88{\tiny\phantom{.}.01}&27.80\\
2010&Dec&13.34&\phantom{0}+46.1&0.091&4.00&27.69{\tiny\phantom{.}.01}&25.60{\tiny\phantom{.}.01}&25.18{\tiny\phantom{.}.00}&24.57{\tiny\phantom{.}.01}&25.28{\tiny\phantom{.}.00}&1.79{\tiny\phantom{.}.05}&1.86{\tiny\phantom{.}.01}&1.85{\tiny\phantom{.}.01}&27.78\\
2010&Dec&13.35&\phantom{0}+46.1&0.091&3.50&27.70{\tiny\phantom{.}.01}&25.60{\tiny\phantom{.}.03}&25.20{\tiny\phantom{.}.01}&24.55{\tiny\phantom{.}.01}&25.23{\tiny\phantom{.}.01}&1.65{\tiny\phantom{.}.06}&1.86{\tiny\phantom{.}.01}&1.85{\tiny\phantom{.}.02}&27.79\\
2010&Dec&13.36&\phantom{0}+46.1&0.091&3.79&27.71{\tiny\phantom{.}.01}&25.58{\tiny\phantom{.}.02}&25.20{\tiny\phantom{.}.00}&24.59{\tiny\phantom{.}.01}&25.26{\tiny\phantom{.}.01}&1.74{\tiny\phantom{.}.04}&1.86{\tiny\phantom{.}.01}&1.83{\tiny\phantom{.}.01}&27.80\\
2010&Dec&13.37&\phantom{0}+46.1&0.091&4.09&27.70{\tiny\phantom{.}.00}&25.54{\tiny\phantom{.}.01}&25.17{\tiny\phantom{.}.00}&24.58{\tiny\phantom{.}.01}&25.27{\tiny\phantom{.}.00}&1.83{\tiny\phantom{.}.03}&1.81{\tiny\phantom{.}.01}&1.77{\tiny\phantom{.}.01}&27.79\\
2010&Dec&13.38&\phantom{0}+46.1&0.091&4.21&27.69{\tiny\phantom{.}.00}&25.56{\tiny\phantom{.}.01}&25.16{\tiny\phantom{.}.00}&24.58{\tiny\phantom{.}.01}&25.28{\tiny\phantom{.}.00}&1.86{\tiny\phantom{.}.03}&1.81{\tiny\phantom{.}.01}&1.76{\tiny\phantom{.}.01}&27.77\\
2010&Dec&13.40&\phantom{0}+46.1&0.091&3.90&27.69{\tiny\phantom{.}.01}&25.57{\tiny\phantom{.}.01}&25.18{\tiny\phantom{.}.00}&24.56{\tiny\phantom{.}.01}&25.26{\tiny\phantom{.}.00}&1.81{\tiny\phantom{.}.03}&1.80{\tiny\phantom{.}.01}&1.78{\tiny\phantom{.}.01}&27.78\\
2010&Dec&13.41&\phantom{0}+46.1&0.091&4.00&27.70{\tiny\phantom{.}.01}&25.61{\tiny\phantom{.}.01}&25.19{\tiny\phantom{.}.00}&24.60{\tiny\phantom{.}.01}&25.29{\tiny\phantom{.}.00}&1.76{\tiny\phantom{.}.05}&1.81{\tiny\phantom{.}.02}&1.77{\tiny\phantom{.}.02}&27.79\\
2011&Jan&\phantom{0}5.30&\phantom{0}+69.0&0.148&4.05&27.42{\tiny\phantom{.}.01}&25.25{\tiny\phantom{.}.04}&24.88{\tiny\phantom{.}.01}&24.21{\tiny\phantom{.}.03}&24.99{\tiny\phantom{.}.01}&1.51{\tiny\phantom{.}.11}&1.67{\tiny\phantom{.}.03}&1.67{\tiny\phantom{.}.03}&27.48\\
2011&Jan&\phantom{0}5.31&\phantom{0}+69.0&0.148&3.84&27.45{\tiny\phantom{.}.02}&25.22{\tiny\phantom{.}.06}&24.85{\tiny\phantom{.}.01}&24.22{\tiny\phantom{.}.03}&24.95{\tiny\phantom{.}.02}&1.70{\tiny\phantom{.}.08}&1.67{\tiny\phantom{.}.03}&1.81{\tiny\phantom{.}.02}&27.51\\
2011&Jan&\phantom{0}5.36&\phantom{0}+69.1&0.148&4.15&27.45{\tiny\phantom{.}.01}&25.23{\tiny\phantom{.}.03}&24.87{\tiny\phantom{.}.01}&24.19{\tiny\phantom{.}.02}&25.01{\tiny\phantom{.}.01}&1.75{\tiny\phantom{.}.06}&1.74{\tiny\phantom{.}.03}&1.73{\tiny\phantom{.}.03}&27.51\\
2011&Jan&\phantom{0}5.37&\phantom{0}+69.1&0.148&4.24&27.44{\tiny\phantom{.}.01}&25.23{\tiny\phantom{.}.03}&24.88{\tiny\phantom{.}.01}&24.24{\tiny\phantom{.}.02}&24.99{\tiny\phantom{.}.01}&1.63{\tiny\phantom{.}.08}&1.68{\tiny\phantom{.}.03}&1.68{\tiny\phantom{.}.03}&27.50\\
2011&Feb&\phantom{0}1.13&\phantom{0}+95.9&0.214&4.23&27.17{\tiny\phantom{.}.04}&24.77{\tiny\phantom{.}.17}&24.74{\tiny\phantom{.}.02}&24.02{\tiny\phantom{.}.07}&24.81{\tiny\phantom{.}.03}&1.59{\tiny\phantom{.}.17}&1.59{\tiny\phantom{.}.07}&1.48{\tiny\phantom{.}.09}&27.20\\
2011&Feb&\phantom{0}1.14&\phantom{0}+95.9&0.214&4.42&27.21{\tiny\phantom{.}.02}&25.08{\tiny\phantom{.}.07}&24.76{\tiny\phantom{.}.01}&24.15{\tiny\phantom{.}.05}&24.79{\tiny\phantom{.}.02}&1.67{\tiny\phantom{.}.14}&1.74{\tiny\phantom{.}.05}&1.82{\tiny\phantom{.}.04}&27.24\\
2011&Feb&\phantom{0}1.21&\phantom{0}+95.9&0.214&4.42&27.19{\tiny\phantom{.}.01}&24.97{\tiny\phantom{.}.07}&24.73{\tiny\phantom{.}.01}&24.14{\tiny\phantom{.}.04}&24.85{\tiny\phantom{.}.02}&1.61{\tiny\phantom{.}.12}&1.58{\tiny\phantom{.}.05}&1.48{\tiny\phantom{.}.07}&27.21\\
2011&Feb&\phantom{0}2.23&\phantom{0}+97.0&0.216&4.54&27.19{\tiny\phantom{.}.01}&24.78{\tiny\phantom{.}.12}&24.71{\tiny\phantom{.}.02}&24.14{\tiny\phantom{.}.06}&24.80{\tiny\phantom{.}.02}&1.34{\tiny\phantom{.}.27}&1.52{\tiny\phantom{.}.07}&1.53{\tiny\phantom{.}.07}&27.22\\
2011&Feb&\phantom{0}2.26&\phantom{0}+97.0&0.216&4.12&27.16{\tiny\phantom{.}.03}&24.46{\tiny\phantom{.}.29}&24.64{\tiny\phantom{.}.03}&23.95{\tiny\phantom{.}.09}&24.69{\tiny\phantom{.}.04}&1.53{\tiny\phantom{.}.19}&1.69{\tiny\phantom{.}.06}&1.59{\tiny\phantom{.}.07}&27.19\\
2011&Feb&23.11&+117.8&0.263&4.37&26.96{\tiny\phantom{.}.05}&24.86{\tiny\phantom{.}.19}&24.52{\tiny\phantom{.}.04}&24.05{\tiny\phantom{.}.10}&24.60{\tiny\phantom{.}.06}&1.46{\tiny\phantom{.}.28}&1.62{\tiny\phantom{.}.09}&1.51{\tiny\phantom{.}.11}&26.96\\
2011&Feb&23.13&+117.9&0.263&4.46&26.95{\tiny\phantom{.}.04}&24.23{\tiny\phantom{.}.40}&24.56{\tiny\phantom{.}.03}&24.00{\tiny\phantom{.}.09}&24.65{\tiny\phantom{.}.05}&1.66{\tiny\phantom{.}.20}&1.53{\tiny\phantom{.}.11}&1.51{\tiny\phantom{.}.11}&26.95\\
2011&Feb&23.15&+117.9&0.263&4.56&26.99{\tiny\phantom{.}.03}&24.78{\tiny\phantom{.}.16}&24.58{\tiny\phantom{.}.03}&24.01{\tiny\phantom{.}.11}&24.63{\tiny\phantom{.}.04}&1.55{\tiny\phantom{.}.23}&1.84{\tiny\phantom{.}.05}&1.73{\tiny\phantom{.}.07}&27.00\\
2011&Feb&23.18&+117.9&0.263&4.37&26.98{\tiny\phantom{.}.04}&24.66{\tiny\phantom{.}.24}&24.59{\tiny\phantom{.}.04}&24.02{\tiny\phantom{.}.10}&24.54{\tiny\phantom{.}.06}&1.43{\tiny\phantom{.}.28}&1.63{\tiny\phantom{.}.08}&1.81{\tiny\phantom{.}.05}&26.98\\

\enddata
\tablenotetext{a} {Production rates, followed by the upper, i.e. the positive uncertainty. The ``+'' and ``-'' uncertainties are equal as percentages, but unequal in log-space; the ``-'' values 
can be computed.}
\tablenotetext{b} {Continuum filter wavelengths: UV (1991) = 3650 \AA; UV (1997/98 \& 2010/11) = 3448 \AA; blue = 4450 \AA; green (1991) = 4845 \AA; green (1997/98 \& 2010/11) = 5260 \AA.}
\tablenotetext{c} {``und'' stands for ``undefined'' and means the continuum flux was measured but was less than 0.}\label{t:phot_rates}
\end{deluxetable}

\begin{deluxetable}{lccccc}  
\tabletypesize{\scriptsize}
\tablecolumns{6}
\tablewidth{0pt} 
\setlength{\tabcolsep}{0.03in}
\tablecaption{Abundance ratios for 103P/Hartley 2}
\tablehead{   
  \colhead{}&
  \colhead{}&
  \colhead{}&
  \multicolumn{3}{c}{log Production Rate}\rule{0cm}{8pt}\\
  \colhead{}&
  \multicolumn{2}{c}{$r_\mathrm{H}$--dependence}&
  \multicolumn{3}{c}{Ratios (X/OH)\tablenotemark{a}}\\
  \cmidrule(lr){2-3}
  \cmidrule(lr){4-6}
  \colhead{Species}&
  \colhead{Pre-Peri}&
  \colhead{Post-Peri}&
  \colhead{Mean}&
  \colhead{$\sigma$$_\mathrm{mean}$}&
  \colhead{$\sigma$$_\mathrm{data}$}
}
\startdata
OH&$-$4.64$\pm$0.22&$-$3.99$\pm$0.05&\phantom{0$-$}0.00&&\rule{0cm}{8pt}\\
NH&$-$7.08$\pm$0.98&$-$4.92$\pm$0.27&\phantom{0}$-$2.18&+0.01&+0.09\\
CN&$-$3.44$\pm$0.18&$-$3.20$\pm$0.10&\phantom{0}$-$2.55&+0.01&+0.07\\
C$_3$&$-$3.93$\pm$0.25&$-$3.20$\pm$0.13&\phantom{0}$-$3.15&+0.01&+0.07\\
C$_2$&$-$4.10$\pm$0.10&$-$3.45$\pm$0.09&\phantom{0}$-$2.46&+0.01&+0.07\\
UC&$-$2.85$\pm$1.26&$-$1.19$\pm$0.19&$-$25.88&+0.03&+0.23\\
BC&$-$3.55$\pm$0.60&$-$0.99$\pm$0.13&$-$25.81&+0.04&+0.25\\
GC&$-$3.35$\pm$1.25&$-$0.97$\pm$0.17&$-$25.84&+0.04&+0.26\\[1.0ex]
UC\tablenotemark{b}\phantom{0}&$-$3.39$\pm$1.28&$-$1.91$\pm$0.20&$-$25.43&+0.03&+0.21\\
BC\tablenotemark{b}\phantom{0}&$-$4.03$\pm$0.63&$-$1.71$\pm$0.15&$-$25.44&+0.04&+0.22\\
GC\tablenotemark{b}\phantom{0}&$-$3.84$\pm$1.26&$-$1.68$\pm$0.19&$-$25.48&+0.03&+0.21\\
\enddata
\tablenotetext{a} {For the dust continuum, the ratio of $Af$$\rho$ to $Q$(OH) has units of cm s mol$^{-1}$. $\sigma$$_\mathrm{data}$ is the standard deviation and measures the scatter of the data values around the sample mean, while $\sigma$$_\mathrm{mean}$ is the standard deviation of the sampling distribution of the mean, or the standard deviation divided by the square root of the number of cases. Only upper error bars are given; the lower error bars can be derived if desired.}
\tablenotetext{b} {$Af$$\rho$ adjusted to $\theta$ = 0$^\circ$.}
\label{t:table5}
\end{deluxetable}

\clearpage

\renewcommand{\baselinestretch}{0.8}

\end{document}